\documentclass[pra, reprint,superscriptaddress,floatfix
]{revtex4-2}

\usepackage{graphicx}% Include figure files
\usepackage{dcolumn}% Align table columns on decimal point
\usepackage{bm}% bold math
\usepackage[utf8]{inputenc}
\usepackage{amsmath,amssymb}
\usepackage{graphicx}
\usepackage{xcolor}
\usepackage{verbatim}
\usepackage{titlesec}
\usepackage[resetlabels,labeled]{multibib}
\newcommand \ket[1]{\left|#1 \right>}
\newcommand \ketbra[2]{\left|#1 \right>\left<#2 \right|}

\newenvironment{shortTitle}
    {\bfseries Short title:\\
    }
    { 
    }

\newenvironment{teaser}
    {\bfseries Teaser:\\
    }
    { 
    }
    
\begin{document}

	\preprint{APS/123-QED}
	
	\title{ Detecting axion dark matter with Rydberg atoms via induced electric dipole transitions 
 }

		\author{Georg Engelhardt}
	\email{engelhardt@sustech.edu.cn}
	\affiliation{Shenzhen Institute for Quantum Science and Engineering, Southern University of Science and Technology, Shenzhen 518055, China}
	\affiliation{International Quantum Academy-Shenzhen, Shenzhen 518048, China}
	\affiliation{Guangdong Provincial Key Laboratory of Quantum Science and Engineering, Southern University of Science and Technology, Shenzhen, 518055, China}
	
	\author{Amit Bhoonah}
	\email{amit.bhoonah@pitt.edu}
	\affiliation{Department of Physics and Astronomy and IQ Initiative,  University of Pittsburgh, Pittsburgh, PA 15260, USA}

	\author{W. Vincent Liu}
	\email{wvliu@pitt.edu}
	\affiliation{Department of Physics and Astronomy  and IQ Initiative,  University of Pittsburgh, Pittsburgh, PA 15260, USA}
 \affiliation{International Quantum Academy-Shenzhen, Shenzhen 518048, China}

\date{\today}

\begin{abstract}
% Science Advance requirement: a single paragraph (less than 160 words)
{
 Long-standing efforts to detect axions are driven by two compelling prospects, naturally accounting for the absence of charge-conjugation and parity symmetry breaking in quantum chromodynamics, and for the elusive dark matter at ultralight mass scale.  Many experiments use advanced cavity resonator setups to probe the magnetic-field-mediated conversion of axions to photons.  Here, we show how to search for axion matter without relying on such a cavity setup, which opens a new path for the detection of ultralight axions, where cavity based setups are infeasible.
 When applied to Rydberg atoms, which feature particularly large transition dipole elements, this effect promises an outstanding sensitivity for detecting ultralight dark matter. 
 Our estimates show that it can provide laboratory constraints in parameter space that so far had only been probed astrophysically, and cover new unprobed regions of parameter space.  
 The  Rydberg atomic gases offer a flexible and inexpensive experimental platform that can operate at room temperature. We project the sensitivity  by quantizing the axion-modified Maxwell equations to accurately describe atoms and molecules as quantum sensors wherever axion dark matter is present.
}

\end{abstract}

\maketitle

\begin{shortTitle}
     Rydberg atoms as cost-effective axion detectors
\end{shortTitle}\\

\begin{teaser}
    Rigorous quantum treatment of the axion-Maxwell equations opens a new path for the detection of axion dark matter. 
\end{teaser}

\section{\label{sec:level1}Introduction}

There is overwhelming astrophysical and cosmological evidence that approximately 85$\%$ of matter in the universe is in the form of non-luminous \textit{dark matter}. Unfortunately, little is known about its nature beyond its gravitational influence on galactic and cosmological scales. While the search for historically popular models like Weakly Interacting Massive Particles (WIMPs) continues, it is important to expand experimental efforts to other well-motivated candidates. One such class of models, ultralight bosonic dark matter, is particularly favoured because of discrepancies that arise when simulations of structure formation with WIMP-like dark matter are compared to observations on galactic scales \cite{1994Natur.370..629M,2011MNRAS.415L..40B,1993MNRAS.264..201K}. These tensions are somewhat alleviated when the dark matter is modelled not as a WIMP-like particle but as an ultralight boson with de-Broglie wavelength comparable to small-scale galactic structures, which corresponds to a mass of order $10^{-21}~\text{eV}/c^2$. 

\begin{figure*}
	\includegraphics[width=\linewidth]{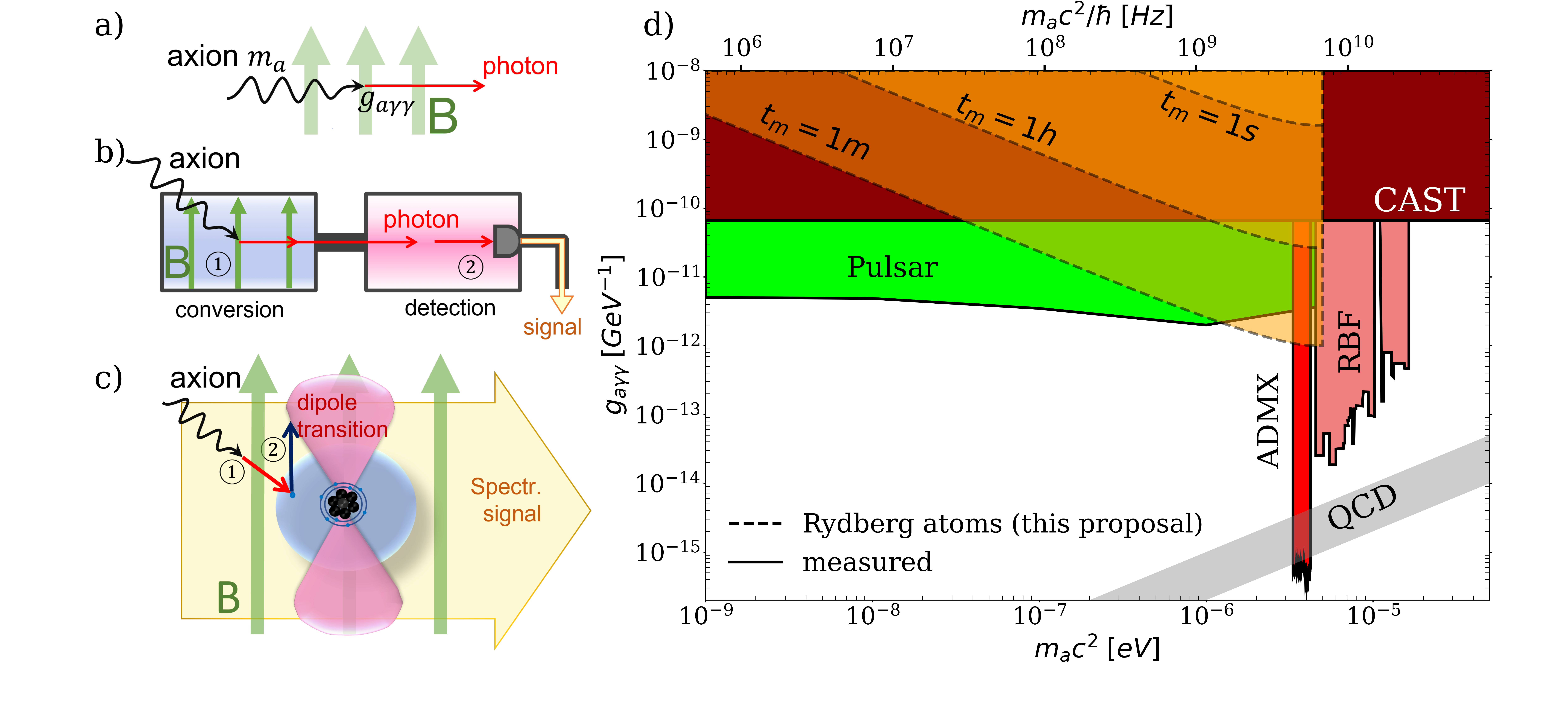}
	\caption{a) Magnetic field meditated axion-photon conversion. b) Typical resonator based setup for the detection of axion dark matter. c) Illustration of the axion-induced electric dipole transition in an atom  mediated by a magnetic field. The dipole transitions can be detected spectroscopically. d) Overview of  exclusion boundaries for  $m_a$ and $g_{a\gamma\gamma}$. Experimentally excluded regions are marked by a solid line, while  proposed experiments are marked by a dashed line. The projected  exclusion bounds using axion-induced dipole transitions in Rydberg atoms achievable within one second, one hour,  and one month measurement time are outlined. Also shown are constraints from: the CERN Axion Solar Telescope (CAST) \cite{CAST:2017uph}, and the  pulsar timing array~\cite{Castillo:2022zfl}.   Other constraints shown on the plot for axions of a higher mass are from ADMX~\cite{Boutan2018} and RBF-UF~\cite{Hagmann1990}.}
	\label{figOverview}
\end{figure*}

The axion is a famous candidate of bosonic dark matter originally predicted as the symmetry consequence of the Peccei-Quinn model as new physics beyond the standard model of elementary particle physics \cite{Peccei:1977hh,Wilczek:1977pj,Weinberg:1977ma,Wilczek:1987mv}. It provides a ``missing" elementary particle capable of naturally explaining why the charge-conjugation and parity symmetry (CP) are preserved in quantum chromodynamics (QCD) but are known to be violated in the electroweak interaction. { This crucial conundrum has been coined the  ``strong CP" problem in elementary particle physics}. While these \textit{QCD axions} may be the dark matter~\cite{Abbott1983,Dine1983,Preskill1983}, it is possible that the latter are pseudo-scalars which, however, do not solve the strong charge-parity problem. To distinguish them from the \textit{QCD axion}, such pseudo-scalars are called axion-like particles, and can usually be searched for with the same tools and techniques used to probe QCD axions. For the remainder of this article, we will use ``axion" to refer to either a dark-matter  candidate  or a new QCD particle whenever the distinction is unimportant. It is interesting to note that, aside from their key role in high energy physics, axions have also been predicted to exist in condensed matter systems \cite{2010NatPh...6..284L,2020NatRP...2..682N,Gooth:2019lmg,Sekine:2020ixs,2016Sci...354.1124W,Li:2014bha} where they manifest in the form of magneto-electric transport effects.  
Thus, a discovery of  the elusive axion particle will not only resolve the strong CP problem in QCD, but could simultaneously  track down the missing dark matter in the universe~\cite{Khlopov1999}. We will focus on the latter aspect in this work.

The axion-modified Maxwell equations predict that a magnetic field converts axions of energy $\hbar \omega_{a} \simeq m_{a}c^2 + \frac{1}{2}m_{a}v^{2}$ (mass $m_a$, velocity $v$)  into photons of frequency $\omega_a$ as sketched in Fig.~\ref{figOverview}(a)~\cite{Wilczek:1987mv}.
A typical experimental setup exploiting this resonant axion-photon conversion for the detection of the galactic axion field consists of two inductively-coupled microwave resonators as depicted in Fig.~\ref{figOverview}(b)~\cite{Kishimoto:1998gv}. A microwave photon, sourced by the axion field in the first resonator, is detected in the second  resonator using a SQUID device, Rydberg atoms, or  comparable single-photon detectors. The RBF-UF~\cite{Hagmann1990},  ADMX~\cite{Boutan2018} and  HAYSTAC~\cite{Brubaker2017,HAYSTAC:2020kwv,PhysRevA.105.042808} experiments attempt to detect the axion field in the narrow 1-200 $\mu$eV mass range to which they are limited by construction. 
It should not be surprising then that most constraints on axions in the mass regime $m_ac^2 <2\mu$eV are astrophysical or cosmological, making alternative laboratory probes valuable.  Current approaches, e.g. ABRACADBRA~\cite{Ouellet2019}, ADMX SLIC~\cite{Crisosto2020} and SHAFT~\cite{Gramolin2021} using lumped-element circuits to compensate for the frequency mismatch lose sensitivity as the mass of the axion is lowered~\cite{Tobar2022,Berlin2021}. Notably a resonant cavity for the detection of ultralight axions would be comparable in size to a small galaxy. Even if possible, the construction of a cavity is complex and cost intensive. 

In this article, we analyze the possibility to search for the ultralight galactic axion field without relying on an advanced cavity resonator setup. Based on a rigorous quantization of the axion-Maxwell equations, we derive an effective Hamiltonian which describes dipole transistion induced by the axion-sourced electric field as sketched in Fig.~\ref{figOverview}(c). Thus, highly sensitive electric field detectors using quantum emitters and spectroscopic methods can be easily repurposed for the search of axion dark matter.

The axion-induced dipole transitions are particularly well suited to Rydberg states which feature long lifetimes, large electric dipole moments and polarizability~\cite{Loew2012}. Being quantum sensors, Rydberg atoms can take advantage of quantized energy levels, quantum coherence, or many-body entanglement to enhance detection sensitivity compared to classical systems~\cite{Degen2017}. Their aforementioned properties make them excellent candidates for electric field metrology in the radio-frequency regime, allowing sensitivities up to $\mu$Vm$^{-1}/\sqrt{\text{Hz}}$~\cite{Jing2020,Holloway2022,Liu2022,Cox2018,Meyer2020}.

After introducing the axion-sourced electric field, we proceed to estimate the  sensitivity of Rydberg atoms in a superheterodyne (superhet) detection configuration. For measurement campaigns over a period between seconds to months we project constraints in the ultralight mass regime that can outperform existing exclusion limits [see Fig.~\ref{figOverview}(d)]. Details of the derivations can be found in the `Methods' section and the Supplementary Materials (SM).

\section{Results}

\subsection{Axion-sourced electric field}

Here, we give a heuristic derivation of the axion-sourced electric field that induces dipole transitions in atoms and similar quantum emitters. A mathematically rigorous derivation is given in the `Methods' section and the SM. The quantization of  the axion-photon Lagrangian $\mathcal{L} = -g_{a\gamma\gamma} \sqrt{\frac{\epsilon_0}{\mu_0}}   a \mathbf{E}\cdot\mathbf{B}$ yields, to linear order in the coupling constant $g_{a\gamma\gamma }$, the axion/light-matter interaction Hamiltonian
\begin{equation}
\hat H_{\text{ALM}} =   g_{a\gamma\gamma }\sqrt{\frac{\epsilon_0}{\mu_0}} \int d^3 \boldsymbol r\, \hat a(\boldsymbol r)    \hat {\boldsymbol E} (\boldsymbol r) \cdot \hat {\boldsymbol B} (\boldsymbol r) ,
\label{eq:axionMatterCoupling_effective}
\end{equation}
where $ \hat {\boldsymbol E} $ and $ \hat {\boldsymbol B} $, denote the electric and magnetic field operators, while $\hat a$ is the (pseudo-scalar) axion field, which interacts via a coupling constant $g_{a\gamma\gamma}$ with $ \hat {\boldsymbol E} $ and $ \hat {\boldsymbol B} $.  $\epsilon_0$ and $\mu_0$ denote the vacuum permittivity and  permeability. The axion field is measured in units of $\text{eV}$ and the interaction constant $g_{a\gamma\gamma}$ in units of $(\text{eV})^{-1}$. 

Crucially,  $ \hat {\boldsymbol E} $ is a physical observable and denotes the total electric field. In the presence of matter, one commonly introduces the displacement field 
\begin{equation}
\hat {\boldsymbol D} (\boldsymbol r)=\epsilon_0\hat {\boldsymbol E} (\boldsymbol r) + \hat {\boldsymbol P} (\boldsymbol r) ,
\label{eq:displacementField0}
\end{equation}
that is sourced by the  density of free charges $\rho_{\text{free}}$ in the electric Gauss equation, i.e., $\boldsymbol \nabla \cdot \hat {\boldsymbol D}   = \rho_{\text{free}}$. The electric field sourced by the bounded charges is described by the polarization density $ \hat {\boldsymbol P}$. 
The displacement field can be understood as the electromagnetic field in the absence of matter and describes thus an external electric field or the quantized electric field in a cavity. The polarization density describes the electric field generated by the matter.

Let us consider an ensemble of  $N$ quantum emitters and  denote their eigenstates   by $\left| i,\mu\right> $, where $i$ labels the quantum emitter, and  $\mu$  represents the collective electronic, vibrational and rotational quantum numbers.
In the dipole approximation,  the  polarization density operator can be expressed in terms of the eigenstates as
\begin{equation}
\hat {\boldsymbol P} (\boldsymbol r)   = \sum_{i=1}^{N}\sum_{\mu,\nu }   \boldsymbol  d^{(i)}_{\mu,\nu}  \left|i, \mu \right> \left< i, \nu \right|   \delta(\boldsymbol r -\boldsymbol r_i),
\label{eq:dipoleDensityOperator0}
\end{equation}
where $\delta(\boldsymbol r)$ is the three-dimensional delta function, $\boldsymbol r_i$ is the position of the $i$-th quantum emitter, and    $\boldsymbol  d^{(i)}_{\mu,\nu}$  is the transition dipole moment  between the two eigenstates $\mu$ and $\nu$. Assuming a stationary magnetic field, we can determine the dynamics of the displacement field sourced by the galactic axion field, which, when interacting with polarization, gives rise to the effective Hamiltonian 

\begin{equation}
	\hat H_{\text{a-d}} = - \sum_{i=1}^{N}   {\boldsymbol E}_{a}(t) \cdot  \boldsymbol  d^{(i)}_{\mu,\nu}  \left|i, \mu \right> \left< i,\nu \right|
	\label{eq:ham:axionDipole}
\end{equation}
and describes dipole transitions driven by the axion-sourced electric field 
\begin{equation}
 {\boldsymbol E}_{a}(t)  = g_{a\gamma\gamma}c\hat  a (t) {\boldsymbol B} \cdot \mathcal S\left(  \frac{m_ac} {\hbar}\cdot R_B  \right)  ,
\label{eq:axionFieldElectricFieldRelation}
\end{equation}
where the function $\mathcal S\left(x  \right)  $ describes the suppression of the axion-sourced electric field  and depends on the axion mass $m_a$ and the length scale $ R_B  $ of the magnetic field.  For small arguments, it scales with $ \mathcal S\left( x\right) \propto x^2$, i.e., the electric field is suppressed for small axion masses, constituting the main challenge to detect  ultralight axions. In this work we assume $ \mathcal S\left( x\right) = x^2$ and a magnetic field length scale of $R_B = 0.1\,\text{m}$.
Relation~\eqref{eq:axionFieldElectricFieldRelation} implies that one can repurpose sensitive atomic and molecular detectors of electric fields to search for axion dark matter. The  (classical) magnetic field can be considered as an experimental switch controlling the effective interaction strength between the axion field and the quantum emitters. This allows to distinguish between a possible axion signal and background  electric fields.

\subsection{Rydberg atoms as axion detectors}

Ultralight axions in the galactic halo exhibit wave-like behavior and must be treated as a classical time-varying background field~\cite{Guth2015},
$ a(t)  = a_0 \cos\left(\omega_{a}t + \phi_{a}\right)$, where $a_{0}$ is the amplitude of the axion field, $\phi_{a}$ its phase, and $\hbar \omega_{a} \simeq m_{a}c^2 + \frac{1}{2}m_{a}v^{2}$ its energy. Since the field has a small frequency dispersion described by the dark matter velocity distribution (average value 10$^{-3}c$), a coherence time can be estimated to be  $\tau_{C}  =  \frac{2\hbar}{m_{a}c^{2}}10^{6}$, which defines the time scale over which the phase $\phi_a$ can be considered to be constant \cite{supplementaryInformation,Foster:2017hbq,Lisanti:2021vij}. For axions in the ultralight mass regime, this time scale will always be much larger than the measurement time in question. The  amplitude  $a_0$ can be estimated  via the galactic dark-matter energy density
$
\rho  =  (m_a^2 c^4 a_0^2 )/(2 \hbar^3 c^3 ) = 0.3 \cdot 10^{15}\frac{eV}{m^3} 
$.
 Resolving Eq.~\eqref{eq:axionFieldElectricFieldRelation} for $g_{a\gamma\gamma}$, we find that  the sensitivity for the interaction parameter is given by
\begin{equation}
g_{a\gamma\gamma,*}   = \frac{  \left| \boldsymbol E_{*}\right|  }{c\left| \boldsymbol B\right|}   \sqrt{ \frac{ m_a^2c^4} {2 \rho c^3\hbar ^3  }}  ,
\label{eq:sensivityToAxionField}
\end{equation}
which assumes that axions constitute 100$\%$ of the dark matter in the universe.

Equation~\ref {eq:ham:axionDipole} shows that atoms with large transition dipole moments, in particular Rydberg states, couple strongly to the axion field. Using an advanced electromagnetically-induced transparence (EIT) based superhet detection protocol, an electric field of $\left| \boldsymbol E_{*}\right|  = 78\cdot 10^{-9} \text{Vm}^{-1} $ could be detected within  $5000$ s measurement time~\cite{Jing2020}.  Taking this setup as the basis for our sensitivity projection, we estimate that Rydberg atoms are able to detect minimal electric fields of $\left| \boldsymbol E_{*}\right| = 30\cdot 10^{-9}\,\text{Vm}^{-1}$, $\left| \boldsymbol E_{*}\right| = 500 \cdot 10^{-12}\,\text{Vm}^{-1} $, and $\left| \boldsymbol E_{*}\right| = 18 \cdot 10^{-12}\,\text{Vm}^{-1}$ within a measurement time of $t_{\text{m}}= 1\,\text{s}$, $t_{\text{m}}= 1\,\text{h}$ and $t_{\text{m}}= 1\,\text{month}$, respectively, under ideal conditions.

Taking these values for reference, Fig.~\ref{figOverview}(d) shows the projected minimal detectable $g_{a\gamma\gamma,*} $ evaluated via Eq.~\eqref{eq:sensivityToAxionField} for a magnetic field of $\left| \boldsymbol B\right| =5.6 \,\text{T}$. It can be readily seen that Rydberg atoms can compete with the CAST helioscope bounds \cite{Adair:2022rtw} in the mass regime $m_a c^2 = 5\cdot 10^{-8} - 5\cdot 10^{-6}\,\text{eV}$. Also shown on the same plot are constraints  inferred from the pulsar timing array \cite{Castillo:2022zfl}.  Rydberg atoms can thus set new leading experimental constrains while being operationally simple to realize. Due to their level structure, Rydberg atoms are particular sensitive in the small frequency regime. Their electric-field sensitivity is thereby relatively independent of the frequency for $\omega< 10\,\text{GHz}$. The exclusion bound established by the Rydberg atoms scales thus with $g_{a\gamma\gamma } \propto m_a^{-1}$, as the suppression factor in Eq.~\eqref{eq:axionFieldElectricFieldRelation} scales with $\propto m_a^2$ and the axon field amplitude with $\propto m_a^{-1}$. For larger masses $m_a c^2 > 5\cdot 10^{-6}\,\text{eV}$, Rydberg atoms perform significantly inferior than the ADMX~\cite{Boutan2018}, and RBF-UF~\cite{Hagmann1990} experiments, whose frequency regimes allows to amplify the axion-sourced electric field in a resonator setup prior to detection.

\subsection{Level structure of Rydberg atoms}

\begin{figure*}
	\includegraphics[width=\linewidth]{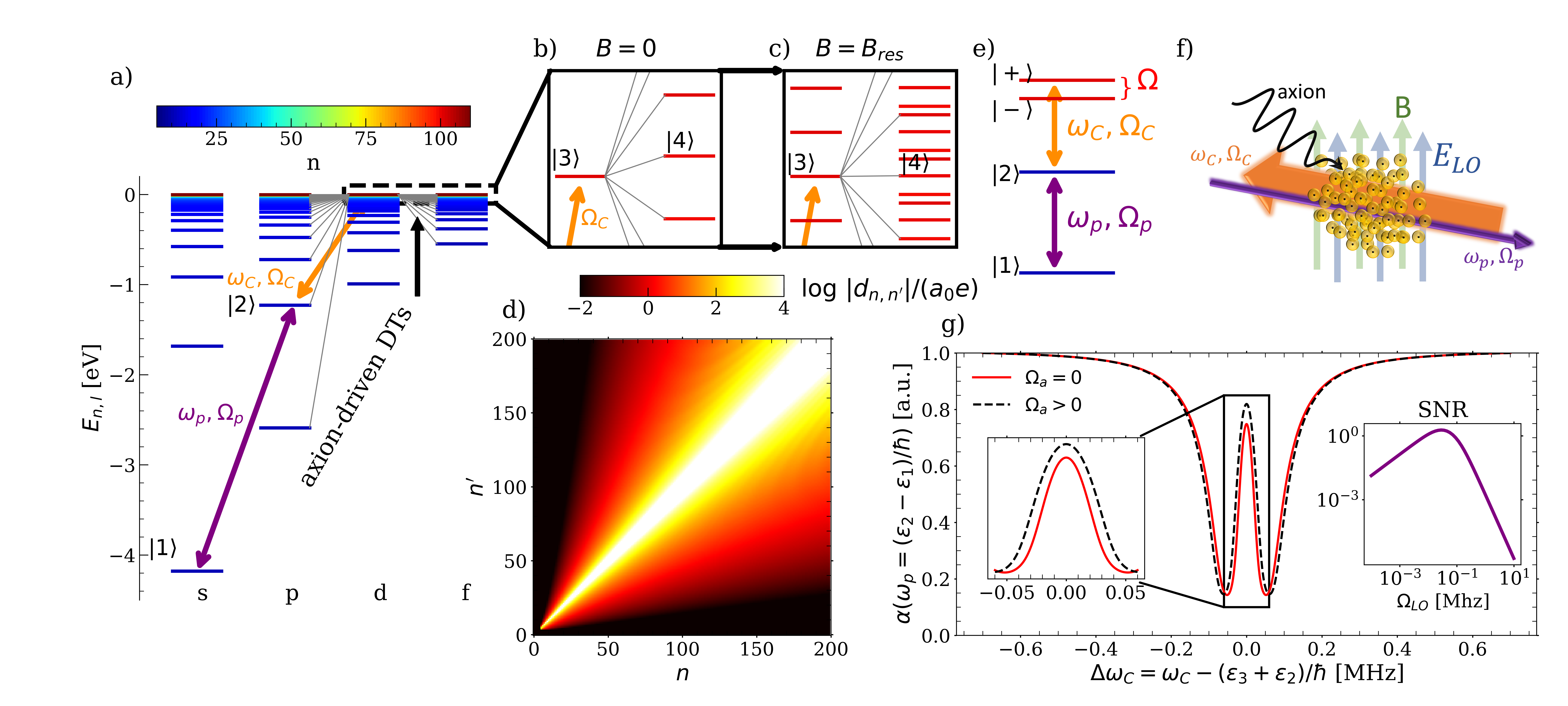}
	\caption{a) Energy levels of a Rubidium atom for $n\geq5$ resolved for the angular momentum orbitals $s,p,d,f$. (b) depicts a magnification of highly-excited $f$ orbitals, that are coupled to the  Rydberg level $\left|3 \right>$ with quantum numbers $n = 100$ and $l=2$ (d orbital). (c) is the same as (b), but for a finite external magnetic field $B=B_{\text{\text{res}}}$, which establishes a resonance condition between levels $\left|3\right>$ with $(n,l,j,m) = (100,2,3/2,1/2)  $  and $\left|4 \right>$  with $(n,l,j,m) = (99,3,5/2,1/2)  $. Gray lines depict the allowed dipole transitions for a linearly-polarized driving field. (d) Transition dipole matrix elements between the $d$ and the $f$ orbitals. (e) Effective four-level system considered in the description of the Rydberg-atom superheterodyne detector. Levels $\left|- \right>$ and $\left|+ \right>$ appear upon coupling the states $\ket{3}$ and $\ket{4}$. (f) Sketch of the spectroscopic setup, in which the probe and coupling fields propagate in the opposite direction to mitigate the Doppler effect as explained in the SM~\cite{supplementaryInformation}. (g) Absorption rate $\alpha(\omega_p)$ of the probe beam as a function of $\Delta\omega_C  = \omega_C -(\epsilon_{3} +\epsilon_2)/\hbar $ at resonance $\hbar\omega_p =\epsilon_2 -\epsilon_1$. The left inset depicts the absorption rate for small $\left|\Delta\omega_C \right| \lesssim \gamma$. The right inset depicts the signal-to-noise ratio (SNR) as a function of the local oscillator $\Omega_{LO}$ at $\Delta \omega_C=0$. Other parameters are $\Omega_p =  1.7\,\text{MHz}$, $\Omega_C = 23\,\text{MHz}$, $B_{\text{res}} =5.6\,\text{T} $, temperature $T=300K$. }
	\label{figExperimentalSetup}
\end{figure*}

Rydberg atoms are highly excited atoms featuring large dipole moments and polarizability.  Each eigenstate can be characterized by the quantum numbers $n, l , j,m$.  The energies $\epsilon_\mu$ depend mainly on the principle quantum number $n\in \left\lbrace 1,2,3,\dots\right\rbrace$ and the angular-momentum quantum number $l = 0, \dots, n-1$ (also denoted by $s,p,d,f\dots)$. The numbers $j$ and $m$ characterize the total angular momentum quantum number and its $z$ projection, respectively. The spectrum of a Rubidium atom is depicted in Fig.~\ref{figOverview}(a), where each column represents a different $l$. The energy splitting between the highly excited Rydberg states $n\gg 1$ rapidly decreases as $n$ becomes large. 
In the presence of a static macroscopic magnetic field $\boldsymbol B \parallel \boldsymbol e_z$, the eigenstates $\left| \mu \right>$ are subject to a Zeemann shift, whose linear contribution reads as
\begin{equation}
	\Delta \epsilon_{\mu} = \frac{\mu_B\left|\boldsymbol B \right|}{\hbar  } \left< \mu \left| \left( \hat L_z + g_z \hat S_z \right) \right|\mu \right> \equiv K_{\mu}\left| \boldsymbol B \right|,
 \label{eq:zeemanShift}
\end{equation}
where $\mu_B$ is the Bohr magneton, $\hat L_z$ is the $z$ projection operator of the angular momentum,  $\hat S_z$ is the projection of the spin, and $g_z$ is the corresponding gyromagnetic factor. The Zeeman effect can be used to tune specific energy states such that the monochromatic axion field fulfills a resonance condition.  

 Because the electron is excited so far away from the nucleus, the transition dipole moments of two neighboring Rydberg states with principle quantum numbers $n^\prime =n\pm1$ are very large and scale as
$  \left| \boldsymbol	d_{\mu,\nu}   \right|^2 \propto n^2$,
given that no optical selection rules are violated, c.f., Fig.~\ref{figExperimentalSetup}(d). Thus, due to their energy spacing and the scaling of their dipole elements, Rydberg atoms are very sensitive to low-frequency and quasistatic electric fields.

\subsection{Sensitivity estimation}

Recent experiments using EIT have demonstrated an outstanding sensitivity of Rydberg-atom  superhet detectors for sensing electric fields~\cite{Jing2020}. 
The superhet configuration consists of two low energy states  $\ket{1}$, $\ket{2}$, and two Rydberg states $\ket{3}$, $\ket{4}$, whose energetic location are marked in Fig.~\ref{figExperimentalSetup}(a) and (b). The states $\ket{1}$ and $\ket{2}$ are coupled by the probe laser of frequency $\omega_p$ and Rabi frequency $\Omega_p$, while the states  $\ket{2}$ and $\ket{3}$ are coupled by the coupling laser of frequency $\omega_C$ and Rabi frequency $\Omega_C$. The Rydberg states are coupled by the axion field of frequency $\omega_a$ and Rabi frequency $\Omega_a =g_{a\gamma\gamma}c  a_0 \left| \boldsymbol d_{3,4} \cdot    \boldsymbol B \right| /\hbar \cdot \mathcal S$. The corresponding energies $\epsilon_{3}$ and $\epsilon_{4}$ are assumed to be close to resonance $\epsilon_{4} - \epsilon_{3}\approx\hbar \omega_a$. The resonance condition can be adjusted using the Zeeman effect [see Fig.~\ref{figExperimentalSetup}(c)]. To enhance the signal, the superhet detector uses a local oscillator with frequency $\hbar \omega_{LO} = \epsilon_{4} - \epsilon_{3}$ and Rabi frequency $\Omega_{LO}\gg\Omega_{a}$ which heterodynes the axion field.

The coupled Rydberg states form the states $\ket{-}$ and $\ket{+}$, which exhibit a slow monochromatic energy splitting $\hbar \Omega(t) =  \epsilon_+(t) - \epsilon_-(t) = \hbar\Omega_{LO} +\hbar \Omega_{a}\cos[ (\omega_a - \omega_{LO})  t] $.  As $\omega_a \approx \omega_{LO}$, we consider $\Omega(t)$ to be quasistatic.   Both states are coupled to state $\ket{2}$ with Rabi frequency $\Omega_C/\sqrt{2}$. The resulting four-level system is depicted in Fig.~\ref{figExperimentalSetup}(e). 

The superhet detector senses the axion field via a change of the absorption of the probe laser traversing a  cloud of Rydberg atoms, as sketched in Fig.~\ref{figExperimentalSetup}(f).  The linear absorption rate $\alpha(\omega_p)$ for  the resonance condition $\hbar \omega_p = \epsilon_2 -\epsilon_1$ is shown in Fig.~\ref{figExperimentalSetup}(g) as a function of $\Delta \omega_C = \omega_C - (\epsilon_{3} +\epsilon_2)/\hbar$. The  absorption rate exhibits a dip around $\Delta \omega_C =0 $ for $\Omega=0$ (not shown). This is the celebrated EIT affect, which is induced by a destructive interference between the transitions from $\ket{1}$ to $\ket{2}$ and $\ket{3}$ ($\ket{4}$) to $\ket{2}$ in the system's stationary state, preventing the absorption of the probe laser~\cite{Scully1997,Fleischhauer2005}.

For a finite  $\Omega$, the atoms are not perfectly transparent closely around  $\left| \Delta\omega_C\right|\approx 0$.
 The energy splitting between the two  states $\left| -\right> $ and $\left| +\right>$  results into two new transparency frequencies at $\Delta \omega_C = \pm \Omega/2$. Their interference shifts the ideal transparency, which would appear for $\Omega=0$ at $\Delta\omega_C=0$, and gives rise to high sensitivity of the absorption rate as highlighted in the left inset of Fig.~\ref{figExperimentalSetup}(g), which compares the absorption rate for $\Omega=\Omega_{LO}$ and $\Omega=\Omega_{LO} +\Omega_a$. A thorough analysis reveals that the signal-to-noise ratio (SNR) for a total measurement time $t_{\text{m} }$ is given by
\begin{equation}
SNR  = \sqrt{ N } \Omega_a \tau , 
\label{eq:projectedSNR0}
\end{equation}
where $N$ is the number of Rydberg atoms,  $\tau = \sqrt{T_c\cdot t_{\text{m} }}$ is the effective measurement time, and $T_c$  is  the effective coherence time
\begin{eqnarray}
T_c &=& \frac{4\Omega_p^2\Omega_C^4\left(  1 +  \Gamma^\prime \right)^2}{\left[ \left( \gamma_2+\frac{\omega_p\sigma}{c}  \Gamma \right)  \left(\Omega_{LO}^2 + \gamma^2  \right)   + \Omega_C^2\gamma  \right]^3}\frac{ \gamma^2  \Omega_{LO}^2 }{\left( \Omega_{LO}^2 +\gamma^2 \right)  }\nonumber  \\
\end{eqnarray}
with $\gamma=(\gamma_3+\gamma_4)/2$,
where $\gamma_2$ , $\gamma_3$, and $\gamma_4$ are the inverse life times of the states $\ket{2}$, $\ket{3}$, and $\ket{4}$, respectively. The Doppler effect for atoms of mass $m_R$ at temperature $T$ is described by  the function $\Gamma = \Gamma\left(\zeta / (\sqrt{2}\sigma)\right)\geq0$, where $\zeta =  \gamma_2+ \gamma\Omega_c^2/\left(\Omega_{LO}^2 + \gamma^2  \right)$ and  $\sigma^2 = k_B T/m_R$, which is defined in the `Methods' section. Its derivative $\Gamma' = \partial_x \Gamma(x)\mid_{x\rightarrow\zeta / (\sqrt{2}\sigma) }$ is bounded by $-0.5<\Gamma^\prime< 0$ and has  a minor impact on $T_c$. The Doppler effect thus enhances the dephasing rate $\gamma_2^\prime \rightarrow \gamma_2 +\frac{\omega_p\sigma}{c}  \Gamma$, which is discussed in detail in the SM.  
Equation \eqref{eq:projectedSNR0} accounts for the projection noise in the atomic system, which sets the only fundamental limit for the SNR, while photon-shot and measurement noises have been neglected here~\cite{Meyer2020}.

As shown in the inset of Fig.~\ref{figExperimentalSetup}(g), the SNR exhibits a turnover as a function of the local oscillator strength. For small $\Omega_{LO}$, it increases linearly with $\Omega_{LO}$ until it reaches a maximum around $\Omega_{LO} \approx 5\gamma$. For  large $\Omega_{LO}$, the SNR vanishes with { $\Omega_{LO}^{-3}$.}  The maximum value of the coherence time is $T_c \approx (\Omega_p/\Omega_C)^2 /\gamma$ which determines the optimal SNR.

Using Eqs.~\eqref{eq:axionFieldElectricFieldRelation} and \eqref{eq:projectedSNR0}, we find an explicit expression for the projected sensitivity of the axion field
\begin{equation} 
g_{a\gamma\gamma,*}   =  \left(\frac{\left| K_{3} -K_{4}\right| }{ \left| \epsilon_{3} -\epsilon_{4}\right|-  m_ac^2 } \right) 
\frac{  \hbar }{\tau\sqrt{N} \left|\boldsymbol d_{3,4} \right|}  \sqrt{ \frac{ m_a^2 } {2 \rho c\hbar ^3  }}  ,
\label{eq:sensivityToAxionFieldRydberg}
\end{equation}
where $K_{3},K_{4}$ are defined in Eq.~\eqref{eq:zeemanShift}.
The first fraction represents the inverse magnetic field $B_{\text{res}}$ required for establishing a resonance condition, which suggests using states featuring similar Zeemann parameters $K_3 \approx K_4$ and a large detuning, as long as the external magnetic field $B_{\text{res}}< 10\,\text{T}$ is experimentally feasible. 
In the proposed Rydberg-atom detector, the magnetic field fulfills thus two tasks: (i) it establishes the resonance condition between the Rydberg states, which enhances the sensitivity to the electric field; (ii) it generates the axion-sourced electric field according to Eq.~\eqref{eq:axionFieldElectricFieldRelation}.  The second fraction represents the minimal electric field which can be detected with the superhet configuration. 

To estimate the projected sensitivity achievable with  Rydberg atoms, we consider the states $\ket{1}= (n=5,l=0,j=1/2,m=1/2)$, $\ket{2}= (n=100,l=1,j=3/2,m=1/2)$, $\ket{3}= (n=100,l=2,j=5/2,m=1/2)$ and $\ket{4}= (n=99,l=3,j=7/2,m=1/2)$ of Rubidium atoms. Their energies and dephasing rates can be calculated using the ARC package~\cite{Robertson2021}. The dipole matrix element of the Rydberg states is $\left| \boldsymbol d_{3,4}\right| = 6425 e a_0$ (Bohr radius $a_0$). In the ultralight axion regime the optimal magnetic field $\left| \boldsymbol B_{\text{res}} \right| \approx 5.6 \text{T}$ determined by the first fraction in Eq.~\eqref{eq:sensivityToAxionFieldRydberg} is almost independent of the axion mass as  $m_ac^2 \ll \epsilon_d -\epsilon_c $. The projected minimal electric fields and the related exclusion limits for $g_{a\gamma\gamma}$ for the measurement times $t_{\text{m}}= 1\,\text{s}$, $t_{\text{m}}= 1\,\text{h}$ and $t_{\text{m}}= 1\,\text{month}$ have been discussed above. 

\section{Discussion}

In this article, we have discussed the possibility to detect the galactic axion field by deploying quantum emitters such as atoms and molecules without using an advanced cavity resonator setup. Our proposal thus facilitates  the search in the ultralight axion regime, where the required cavity length becomes unreasonable long.
Based on a rigorous quantization of the axion-Maxwell equations, we have derived an effective Hamiltonian that microscopically describes dipole transitions in atoms, molecules, trapped ions driven by the axion-sourced electric field.
 This presents an exciting opportunity for re-purposing existing electric field detectors based on atomic quantum sensors, which promise performance enhancement by means of quantum engineering~\cite{Wang2022}, for axion detection. In this article, we proposed one such method using highly excited Rydberg states, whose large transition dipole elements make them excellent probes of axion-induced dipole transitions. We further conjecture that the dipole transitions  can also enhance the sensitivity of other axion detectors, such as helioscopos~\cite{CAST:2017uph} and   phonon polaritons~\cite{Mitridate2020}. We note that the  direct detection scheme proposed here is different from previous protocols using Rydberg atoms, which  either employ the atoms for an indirect detection of the axion-sourced photon in a cavity~\cite{Yamamoto1999}, or the coupling of the electron spin to the axion wind~\cite{Matsuki1991,Sikivie2014,Flambaum2018a}.

It is useful to consider further sensitivity improvement of our setup. The sensitivity may be improved for longer measurement campaigns over several year, similar to the one performed by the BACON collaboration \cite{2020arXiv200514694O} searching for the time variation of fundamental constants. In this case one would also have to carefully account for the stochastic fluctuations of the axion dark matter field. Another avenue might be to seek improvements in the measurement process itself to facilitate sensing of weaker electric fields, e.g., by using stronger probe fields. The sensitivity estimate of the superhet detector for weak probe fields shows that the SNR is proportional to $\Omega_p/\Omega_C$, i.e., it improves for stronger probe fields. However, this requires the development of non-perturbative theoretical methods, e.g., based on the photon-resolved Floquet theory~\cite{Engelhardt2022a}, to accurately predict the spectroscopic signatures of the axion dark matter. One interesting possibility proposed in \cite{Gilmore:2021qqo} is to use trapped ion crystals which can achieve electric field sensitivities of 100 nVm$^{-1}$.  Moreover, trapping the Rydberg atoms in an optical lattice can further help to mitigate the Doppler effect, and would allow for complex quantum operations to mitigate measurement noise~\cite{Degen2017,Bluvstein2022}.

\section{Methods}
\allowdisplaybreaks

\subsection{Quantization of the axion Maxwell equations}

\label{sec:axionMaxwellEquations_quantization}

The axion field, being a pseudo-scalar, interacts with the electric and magnetic fields via the Lagrangian term $\mathcal{L} = -  g_{a\gamma\gamma} \sqrt{\frac{\epsilon_0}{\mu_0}  } a \mathbf{E}\cdot\mathbf{B}$, where $\boldsymbol  E$, $\boldsymbol  B$, and $a$ denote the classical electric, magnetic, and axion fields, respectively. After deriving the Euler-Lagrange equations, we obtain the axion-Maxwell equations~\cite{Wilczek:1987mv,supplementaryInformation}
\begin{eqnarray}
\boldsymbol \nabla\cdot \boldsymbol  E &=& \frac{\rho}{\epsilon_0}  - cg_{a \gamma\gamma } \boldsymbol  B \cdot \boldsymbol \nabla a \label{eq:elelctricGauss} ,\\
\boldsymbol  \nabla\times\boldsymbol   B - \frac{\dot {\boldsymbol  E}}{c^2}    &=& \mu_0  \boldsymbol  J \nonumber \\ 
&+& \frac{ g_{a\gamma\gamma }}{c} \left(\boldsymbol  B  \dot a - \boldsymbol  E \times\boldsymbol  \nabla a \right) \label{eq:ampere} , \\
\boldsymbol  \nabla \cdot \boldsymbol  B  &=&  0   \label{eq:magneticGauss},  \\
\boldsymbol   \nabla \times \boldsymbol  E  + \dot {\boldsymbol  B }  &=&   0  \label{eq:faraday}, \\
 \ddot a - c^2\boldsymbol \nabla^2 a+ \frac{m_a^2c^4} {\hbar^2}  a   &=&  \hbar c^3\sqrt{\frac{\epsilon_0}{\mu_0}  } g_{a\gamma\gamma} \boldsymbol  E\cdot \boldsymbol   B   \label{eq:axionEom},
\end{eqnarray}
where $\boldsymbol  J$ is the electric current density.  The constants have been described in the `Results' section.
Clearly, These equations reduce to the common Maxwell equations for $g_{a\gamma\gamma}=0$.

As we show in  details in the SM, the canonically quantized axion-light-matter Hamiltonian reads as
\begin{eqnarray}
\hat  H&=& \frac 12 \int d^3 \boldsymbol r\, \left[ \epsilon_0\hat{\boldsymbol E}_c^{\perp 2 }  +  \frac{1}{\mu_0}\hat{\boldsymbol B}^2   \right]  \nonumber \\
 &-& \int d^3 \boldsymbol r\, \left[   \sqrt{\frac{\epsilon_0}{\mu_0}  }  g_{a\gamma\gamma}a\hat{\boldsymbol E}_c \cdot \hat{\boldsymbol B} - \frac{1}{2\mu_0} \left(g_{a\gamma\gamma} \hat a\right)^2 \hat{\boldsymbol B}\cdot \hat{\boldsymbol B} \right]  \nonumber \\
&+&\frac{1}{2} \int d^3 \boldsymbol r\, \left[ \frac{\hat \pi^2 }{\hbar c^3}  +  \frac{1}{c\hbar} \boldsymbol\nabla \hat a \cdot  \boldsymbol \nabla \hat  a  + \frac{ m_a^2c^4}{c^3\hbar^3} \hat a^2\right] \nonumber  \\
&+& \sum_{\eta}\left[  \frac{\hat {\boldsymbol p}^2_{\eta }}{2 m_{\eta}}  +  \frac{q_{\eta}^2}{2 m_{\eta} c^2} \hat {\boldsymbol A}^2(\hat{ \boldsymbol r}_{\eta}) \right] \nonumber  \\
&+& \frac{1}{2} d^3 \boldsymbol r\, \left[\hat \rho \hat \phi + \hat {\boldsymbol J} \cdot \hat {\boldsymbol A}  \right]
\label{eq:hamiltonian0},
\end{eqnarray}
where the canonical electric field operator has been distributed in the transversal and longitudinal  fields  $\hat {\boldsymbol E}_c = \hat {\boldsymbol E}^{\perp}_c +\hat {\boldsymbol E}^{\parallel }_c $. The  transversal  contribution is defined by $\boldsymbol \nabla \cdot  \hat {\boldsymbol E}^{\perp}_c  =0$, while the longitudinal contribution is given by  $\hat {\boldsymbol E}^{\parallel }_c  = \hat {\boldsymbol E}_c - \hat {\boldsymbol E}^{\perp}_c = - \boldsymbol \nabla \hat \phi$, where $\hat \phi$ is the electrostatic vector potential . The vector potential and the magnetic field operators are denoted by $\hat {\boldsymbol B}$ and $\hat {\boldsymbol A}$, respectively.  The axion field is denoted by $\hat a$, and $\hat \pi$ denotes its conjugated momentum.  Particles with charge $q_\eta$ and mass $m_\eta $ at positions $\hat {\boldsymbol  r}_\eta$ having momentum $\hat {\boldsymbol  p}_\eta$ are labeled by $\eta \in \left\lbrace 1,\dots, N_p\right\rbrace$.
 The coupling between light and matter reflects the minimal coupling principle.  The relation between the canonical and  the physical (observable)  electric field operators $\hat {\boldsymbol E}$ is established  via the relation
\begin{equation}
\hat {\boldsymbol E}  = 	\hat{\boldsymbol E}_c - cg_{a\gamma\gamma} \hat a \hat{ \boldsymbol  B}.
\label{eq:rel:canonicalPhysical0}
\end{equation}
The canonical operators for all other operators agree with the physical operators. The field operators are quantized as
\begin{eqnarray}
\hat \rho(\boldsymbol  r) &=& \sum_{\eta=1}^{N_p} q_{\eta} \delta \left({\boldsymbol  r} - \hat{ \boldsymbol  r}_{\eta} \right) ,\\
\hat {\boldsymbol  J }(\boldsymbol  r) &=& \sum_{\eta=1}^{N_p} q_{\eta}\dot{ \hat{ \boldsymbol {r}}}_{\eta} \delta \left( {\boldsymbol  r} - \hat{ \boldsymbol  r}_{\eta}  \right)+ \text{h.c.} , \\
\hat {\boldsymbol A} (\boldsymbol r) &=& \sum_{\boldsymbol k,\lambda} {\boldsymbol e}_{\boldsymbol k,\lambda} \sqrt{ \frac{\hbar }{2\omega_{\boldsymbol k}\epsilon_0 V}} \left( \hat d_{\boldsymbol k ,\lambda}^\dagger  e^{i\boldsymbol k\cdot \boldsymbol r}  +  \hat d_{\boldsymbol k ,\lambda}  e^{-i\boldsymbol k\cdot \boldsymbol r}   \right),  \label{eq:def:vectorPotential} \nonumber \\   \\
\hat{\boldsymbol E}_c^{\perp }(\boldsymbol r) 
&=&  i\sum_{\boldsymbol k,\lambda} {\boldsymbol e}_{\boldsymbol k,\lambda} \sqrt{ \frac{\hbar\omega_{\boldsymbol k}   }{2 \epsilon_0V}} \left( \hat d_{\boldsymbol k ,\lambda}^\dagger  e^{i\boldsymbol k\cdot \boldsymbol r}  -  \hat d_{\boldsymbol k ,\lambda}  e^{-i\boldsymbol k\cdot \boldsymbol r}   \right) , \label{eq:def:electricFieldPerp}   \nonumber \\  \\
\hat{\boldsymbol E}_c^{\parallel}( \boldsymbol r)  &=& -\boldsymbol \nabla \hat \phi( \boldsymbol r)  \label{eq:def:electricFieldPara}, \nonumber \\
 \hat \phi( \boldsymbol r)  &=&  \frac{1}{4\pi\epsilon_0}\sum_{\eta=1}^{N_p}\frac{q_{\eta}}{\left| \boldsymbol r - \hat{\boldsymbol r}_{\eta} \right|},\\
\hat {\boldsymbol B}(\boldsymbol r) &=&   i\sum_{\boldsymbol k,\lambda} \sqrt{ \frac{\hbar  }{2\omega_{\boldsymbol k}  \epsilon_0 V}}  \left( \boldsymbol k  \times {\boldsymbol e}_{\boldsymbol k,\lambda}   \right)  \nonumber \\
&&\times \left( \hat d_{\boldsymbol k ,\lambda}^\dagger   e^{i\boldsymbol k\cdot \boldsymbol r}  -  \hat d_{\boldsymbol k ,\lambda}   e^{-i\boldsymbol k\cdot \boldsymbol r}   \right).
 \label{eq:def:magneticField}
\end{eqnarray}
The photonic operators $\hat d_{\boldsymbol k,\lambda}$  quantizing the electromagnetic field are labeled by wavevector $\boldsymbol k$ and polarization $\lambda \in\left\lbrace  \updownarrow, \leftrightarrow \right\rbrace$. The frequencies of the photonic modes are given by $\omega_{\boldsymbol k}$.  The quantization volume is denoted by $V$. The unit vectors $ {\boldsymbol e}_{\boldsymbol k,\lambda}$ describe the direction of polarization and are perpendicular to the wavevector, i.e., ${\boldsymbol e}_{\boldsymbol k,\lambda} \cdot \boldsymbol k =0 $. This fact readily ensures that the electric and magnetic Gauss equations in Eqs.~\eqref{eq:elelctricGauss} and \eqref{eq:magneticGauss}  are fulfilled. Using the Heisenberg equations of motion and the relation between canonical and physical electric fields in Eq.~\eqref{eq:rel:canonicalPhysical0}, we can derive the axion-Maxwell equations Eq.~\eqref{eq:elelctricGauss}-\eqref{eq:axionEom} in the classical limit by generalizing the common derivation of the Maxwell equations~\cite{Mukamel1995}.

\subsection{Multi-polar Hamiltonian}

The minimal-coupling Hamiltonian in Eq.~\eqref{eq:hamiltonian0} is inconvenient to deal with because of the vector potential  $ \hat{ \boldsymbol A}$ that appears quadratically. This causes more complicated expressions when calculating, e.g., the spectroscopic response of quantum systems. Moreover, the vector potential $\hat{\boldsymbol A}$ is not a physical observable. For this reason, we will bring the Hamiltonian into the so-called multi-polar form that contains only the electric and magnetic fields~\cite{Mukamel1995}. This is done by means of the Power-Zienau transformation~\cite{Power1957}, which we review here shortly. More details can be found in the SM~\cite{supplementaryInformation}. 
Here, we restrict our investigation to systems with bounded charges, such as atoms, molecules, and similar quantum emitters. We will specify to atoms in the following for clarity. Accordingly, the summation over $\eta$ in Eq.~\eqref{eq:hamiltonian0} will be modified into a double summation over $i \in \left\lbrace1,\dots,N \right\rbrace$, labeling atoms,  and $j \in \left\lbrace 1,\dots,N_c \right\rbrace$, labeling the charges associated with a particular atom.  The center of mass of atom $i$ will be denoted by $\boldsymbol R_i$, while the position of the charges belonging to this atom is denoted by $\boldsymbol r_{i,j}$.

The Power-Zienau transformation  is defined  by the unitary operator
\begin{equation}
\hat U  = \exp\left[ -i\frac{1}{\hbar} \int d \boldsymbol r^3 \hat {\boldsymbol P} (\boldsymbol r)  \cdot \hat {\boldsymbol A} (\boldsymbol r)\right],
\end{equation}
where
\begin{eqnarray}
\hat {\boldsymbol P} (\boldsymbol r)  &=&   \sum_{m,j}\hat {\boldsymbol n}_{i,j}(\boldsymbol r),\nonumber \\ 
\hat {\boldsymbol n}_{i,j}(\boldsymbol r)  &=&  q_{i,j} \left( \hat {\boldsymbol r}_{i, j} - \boldsymbol R_{i} \right) \nonumber \\
 &\times & \int_{0}^{1} du  \;\delta \left[ \boldsymbol r - \boldsymbol R_i - u \left( \hat{\boldsymbol r}_{i, j}  - \boldsymbol R_{i} \right) \right]
\end{eqnarray}
denote the macroscopic polarization and the polarization generated by the charges $\eta=(i,j)$, respectively. For simplicity, we assume that there are no free charges in the system, such that $\hat \phi =0$.  

The Power-Zienau transformation leaves all operators in Eq.~\eqref{eq:hamiltonian0} invariant, except of the canonical electric field, which becomes
\begin{equation}
\hat {\boldsymbol D} (\boldsymbol r) \equiv    \hat U \hat {\boldsymbol E}_c (\boldsymbol r)  \hat U^\dagger   =  \hat {\boldsymbol E}_c (\boldsymbol r)   + \frac{1}{\epsilon_0}\hat  {\boldsymbol P} ,
\end{equation}
 and is called the displacement field $ \hat {\boldsymbol D} (\boldsymbol r)$. Importantly,  the displacement field is quantized in terms of the  photonic operators $\hat d_{\boldsymbol k,\lambda}$:
\begin{eqnarray}
\hat {\boldsymbol D}^\perp (\boldsymbol r)  =  i \sum_{\boldsymbol k,\lambda} \sqrt{ \frac{\hbar\omega_{\boldsymbol k}\epsilon_0   }{2 V }}    \boldsymbol  e_{k,\lambda} \left[\hat d_{\boldsymbol k,\lambda}    e^{i\boldsymbol k\cdot \boldsymbol  r} - \hat d_{\boldsymbol k,\lambda}^{\dagger}   e^{-i \boldsymbol k\cdot\boldsymbol r}   \right] ,\nonumber\\
\label{eq:quantized_displacementField}
\end{eqnarray}
and not the canonical electric field $\hat {\boldsymbol E}_c (\boldsymbol r) $.  However, one should keep in mind that the electric field $\hat {\boldsymbol E} $ is the actual physical observable.   Away from the matter where $ \hat {\boldsymbol P} (\boldsymbol r) =0$, the displacement field is equivalent to the canonical electric field $\hat {\boldsymbol D} (\boldsymbol r)=\epsilon_0\hat {\boldsymbol E}_c (\boldsymbol r)$. 

The transformed axion-light-matter coupling Hamiltonian ($\propto a \hat {\boldsymbol E}_c \cdot \hat {\boldsymbol B}$) now reads as 
\begin{eqnarray}
\hat H_{\text{Int}} &=& c g_{a\gamma\gamma} \int d \boldsymbol r^3  \; \left[ \hat a \left( \hat {\boldsymbol D}   -  \hat {\boldsymbol P}  \right)  \cdot \hat  {\boldsymbol B}\right] \nonumber \\
 &+& \int d \boldsymbol r^3  \; \left[  \frac{1}{2\mu_0} g_{a\gamma\gamma}^2 \hat a^2 \hat{\boldsymbol B}\cdot \hat{\boldsymbol B} \right]
\label{eq:SM:axionCoupling}, 
\end{eqnarray}
where we have used that $c^2 = 1/ (\epsilon_0 \mu_0) $. The first term is equivalent to $H_{\text{ALM}}$ in Eq.~\eqref{eq:axionMatterCoupling_effective} after substituting $\hat {\boldsymbol E}$ according to Eq.~\eqref{eq:displacementField0}. As the second term is proportional to $g_{a\gamma\gamma  }^2$, it can be safely neglected. The other terms in the Hamiltonian in Eq.~\eqref{eq:hamiltonian0} transform according to the common Power-Zienau transformation and are given in the SM~\cite{supplementaryInformation}.

\subsection{Rydberg atoms as superheterodyne detectors}

At present, there is a variety of electric field detectors that deploy an EIT coupling scheme in Rydberg atoms. The common EIT three-level configuration is easy to realize experimentally and allows already for excellent sensitivity to weak electric fields~\cite{Holloway2022,Liu2022,Cox2018,Meyer2020}. Yet, in a three-level configuration, the signal field does only perturbatively couple off-resonant Rydberg states to each other, which limits the optimal sensitivity. To further improve the sensitivity, the superhet detection scheme has been developed, which takes advantage of a resonance condition of two Rydberg states, that features a large transition dipole matrix element~\cite{Jing2020}.

A superhet detector deploying Rydberg states can be described by an effective four-level system whose states are denoted by $\left| 1\right>$, $\left| 2 \right>$,  $\left| 3\right>$, $\left| 4 \right>$. Their positions in the energy spectrum are shown in Fig.~\ref{figExperimentalSetup}. The corresponding Hamiltonian  reads as
\begin{eqnarray}
H(t) 
 &=& \epsilon_{1} \ketbra{1}{1}  +  \epsilon_{2} \ketbra{2}{2}  +  \epsilon_{3} \ketbra{3}{3}  + \epsilon_{4} \ketbra{4}{4}  \nonumber \\
&+& \left[  \hbar \Omega_{p} e^{i\omega_p t}  \ketbra{2}{1} + h.c.  \right] \nonumber\\
&+& \left[\hbar\Omega_{C} e^{i\omega_C t} \ketbra{3}{2} + h.c.  \right]\nonumber  \\
&+& \left[\hbar \Omega(t) e^{i\omega_{LO} t }   \ketbra{4}{3} + h.c. \right],
\label{eq:ham:superhet}
\end{eqnarray}
where $\epsilon_{x}$ with $x=1,2,3,4$ denote the level energies. A description of the parameters $\omega_p$, $\omega_C$, $\omega_a$, $\omega_{LO}$, $\Omega_p$, $\Omega_C$, $\Omega_a$, $\Omega_{LO}$ is given in the `Results' section. We consider the time-dependent Rabi frequency $\Omega(t) = \Omega_{LO}  +  \Omega_a e^{i( \omega_a -\omega_{LO} ) t } $ to be quasistatic, and use the difference of the measurement signal for $\Omega = \Omega_{LO} - \Omega_{a} $ and $\Omega =\Omega_{LO} + \Omega_{a}$ to estimate the SNR.
We describe the dynamics of the four-level system in the presence of dissipation  by the Bloch equation
\begin{eqnarray}
\frac{d}{dt}\rho &=& -\frac{i}{\hbar} \left[ H(t) ,\rho \right] \nonumber \\
&+& \gamma_2 D_ { \ketbra{1}{2}} \rho   + \gamma_3 D_ { \ketbra{1}{3}}\rho  + \gamma_4 D_ { \ketbra{1}{4}}    \rho, 
\label{eq:BlochEquation}
\end{eqnarray}
where  the dissipator is defined by
\begin{equation}
D_{\hat O } \rho = 2 \hat O \rho \hat O^\dagger  - \hat O^\dagger \hat O \rho - \rho \hat O^\dagger \hat O ,
\end{equation}
and $\gamma_2$ , $\gamma_{3} $ and $\gamma_{4}$ denote the inverse lifetimes of the states  $\left| 2 \right>$,  $\left| 3\right>$, $\left| 4 \right>$, respectively.

Transforming the subsystem of the states $\left| 3\right>$, $\left| 4 \right>$ into a frame rotating with $\omega_{LO}$, the effective energies of the resulting mixed Rydberg states become $\epsilon_{-} = \epsilon_{3} -\hbar  \Omega/2$ and $\epsilon_{+} = \epsilon_{3} + \hbar\Omega/2$, which are sketched in Fig.~\ref{figExperimentalSetup}(e).  Both mixed Rydberg levels couple to state $\ket{2}$ with Rabi frequency $\Omega_C/\sqrt{2}$ and have a dephasing rate of $\gamma =\frac{1}{2} \left( \gamma_{3} +\gamma_{4}\right)$.

As common practice in EIT experiments, we aim to detect the presence of the axion field by measuring the absorption of the probe field. When certain resonance conditions are fulfilled by the probe and coupling fields, the atoms become transparent to the probe laser within a very narrow window around $\omega_p = (\epsilon_2 -\epsilon_1)/\hbar$. As the resonance conditions are  modified in the presence of the axion field, the absorption significantly depends on $\Omega_a$, heralding the presence of axion dark matter.

The  absorption rate $\alpha(\omega_p)$ of the probe beam in Fig.~\ref{figExperimentalSetup}(f) is proportional to the  susceptibility $\alpha(\omega_p) \propto \text{Im}\, \chi(\omega_p)  $~\cite{Mukamel1995}. As explicitly shown in the SM, the susceptibility can be expressed in terms of the operator $\hat {\mathcal X}  =  \sum_{i=1}^{N} \hat {\mathcal X}_i $ with
\begin{eqnarray}
\hat {\mathcal X}_i =\int_{0}^{t_{\text{m}} } dt \;\hat {\mathcal P}_{2,1 }^{(i)} (t) i e^{ i\omega_p t } + h.c., \nonumber  \\
\hat {\mathcal P}_{2,1 }^{(i)} (t) \equiv    \hat U^\dagger(t)  \left( \left| 2 \right>_i\left< 1\right| +\left| 1 \right>_i\left< 2\right|  \right)\hat U(t) ,
\end{eqnarray}
where the subscript $i$ labels the $N$ atoms in the ensemble and $t_{\text{m}}$ denotes the measurement time. As  $\hat U(t) $ is the time-evolution operator associated with the Bloch equation in Eq.~\eqref{eq:BlochEquation},  $\hat {\mathcal X}$ acts nonlocal in time. 

Explicitly, the relation between the expectation value of $ \hat  {\mathcal X} $ and the susceptibility  in linear order of $\Omega_p$ is given by
\begin{eqnarray}
\text{Im} \,\chi(\omega_p)   & =  &   \frac{1}{2\Omega_p    t_{\text{m}} N}      \frac{\rho_N \left|\boldsymbol d_{1,2}\right|^{2}  }{\epsilon_0\hbar } \left< \hat {\mathcal X}\right>  ,  
 \label{eq:signalSusceptibiliyRelation0}
\end{eqnarray}
where $\rho_N$ denotes the atom density.  The expectation value $\left< \bullet\right> =\text{tr} \left[\bullet \rho_{\text{st}}  \right]$  is defined in terms of the stationary state of the system for $\Omega_p =0$ that reads $\rho_{\text{st}}  = \ketbra{1}{1}$. For this reason,   $\hat {\mathcal X} $ can be considered as a susceptibility operator, and be employed to estimate the SNR.  A straightforward perturbative treatment in $\Omega_p$, which  generalizes the standard treatment of the EIT~\cite{Scully1997} to the four-level system under investigation, and a velocity average to account for the Doppler effect,  reveals that the susceptibility is given by
\begin{eqnarray}
\text{Im } \chi(\omega_p)  &=&      \frac{\rho_N  \left|\boldsymbol d_{1,2}\right|^2}  {\epsilon_0\hbar\Omega_p }\text{Im }\; \frac{c_{1}}{c_{2} + i \sigma  \Gamma(-ic_{2}/\sigma) },   
\label{eq:susceptibilityFrequencyDoppler}
\end{eqnarray}
where we have defined
\begin{eqnarray}
c_{1} &=& -\frac{\lambda_p\Omega_p}{2\pi  }, \nonumber  \\
c_{2} &=& \frac{\lambda_p} {2\pi}\left(   \tilde \epsilon_{12}  - \frac{1}{8} \frac{\Omega_{C}^2}{ \tilde \epsilon_{1-} }- \frac{1}{8} \frac{\Omega_{C}^2}{  \tilde \epsilon_{1+} } \right) \nonumber , \\
\sigma &=&\sqrt{ \frac{k_B T}{m_R}}, \nonumber \\
 \Gamma(z)  &=&  \sqrt{\frac{2}{\pi}} \frac{e^{-z^2}}{1 - \text{erf} \left( z\right)    } -\sqrt{2}z,
\end{eqnarray}
where $\text{erf} \left( z\right)$ denotes the complex-valued error function,  and we have introduced
\begin{eqnarray}
\tilde \epsilon_{12}   &=& (\epsilon_1 -\epsilon_2)/\hbar +\Omega_p + i\gamma_2 \nonumber, \\
\tilde \epsilon_{1-}  &=& (\epsilon_1 -\epsilon_{3})/\hbar  - \frac{\Omega}{2} +\omega_C + \omega_p  + i\gamma, \nonumber  \\
\tilde \epsilon_{1+}   &=&  (\epsilon_1 -\epsilon_{4})/\hbar  + \frac{\Omega}{2}+\omega_C + \omega_p  + i\gamma ,
\label{eq:parameterDefinitions}
\end{eqnarray}
for a rotational reason.

The absorption rate (i.e., the imaginary part of the susceptibility) is depicted in Fig.~\ref{figExperimentalSetup}(g) as a function of $\Delta \omega_{C} =\omega_C -(\epsilon_{3} + \epsilon_1)/\hbar  $ for $\omega_p = (\epsilon_2 -\epsilon_1)/\hbar $, for a suitable local oscillator strength $\Omega_{LO}$,  and  two different $\Omega_a =0 $ (dashed) and $\Omega_a >0 $ (solid).  In the inset we depict the susceptibility for frequencies close to $\left| \Delta \omega_C  \right| \lesssim \gamma$, where we observe a clear difference between the cases  $\Omega_a =0 $  and $\Omega_a >0 $.
For this reason, we choose the coupling frequency to be  $\Delta \omega_{C}=0$, for which the signal for small $\Omega_a$ is given by
\begin{eqnarray}
\delta  \left< \hat {\mathcal X}\right>   &\equiv&  \Omega_a \left.\frac{d}{d\Omega_a} \left< \hat {\mathcal X}\right> \right|_{\Omega_a =0}  \nonumber \\
&=&           \frac{  4\Omega_C^2\gamma   \Omega_{LO}\Omega_p   \cdot N  \Omega_a t_{\text{m}}  \left(  1 +  \Gamma^\prime \right)  }{ \left[ \left( \gamma_2+\frac{\omega_p\sigma}{c}  \Gamma  \right)  \left(\Omega_{LO}^2 + \gamma^2  \right)   + \Omega_C^2\gamma \right]^2    }    , 
\label{eq:superhero:signal}\\
\nonumber
\end{eqnarray}
where we have defined $\Gamma =  \Gamma\left(\zeta / (\sqrt{2}\sigma)\right)$ with $\zeta =  \gamma_2+ \gamma\Omega_c^2/\left(\Omega_{LO}^2 + \gamma^2  \right)$, and $\Gamma' = \partial_z  \Gamma(z)\mid_{z\rightarrow \zeta / (\sqrt{2}\sigma) }$.
Likewise, we find that the variance of $\hat {\mathcal X}$  for vanishing $\Omega_a$ is given as
\begin{eqnarray}
\text{Var}\left< \hat {\mathcal X} \right>  &=&	   t_{\text{m}} N  \, \frac{\epsilon_0\hbar } {\rho_N\left|\boldsymbol d_{1,2}\right|^{2}  } \text{Im}\, \chi(\omega_p) \nonumber  \\
&=&	     \frac{4 t_{\text{m}} N }{\left( \gamma_2+\frac{\omega_p\sigma}{c}  \Gamma  \right)  + \frac{\Omega_C^2\gamma  }{\Omega_{LO}^2 + \gamma^2 }     } , 
\end{eqnarray} 
where we have approximated $\Omega\rightarrow \Omega_{L0}$. Putting everything together, we obtain the $SNR \equiv \delta  \left< \hat {\mathcal X}\right>  /\left(  \text{Var}\left< \hat {\mathcal X} \right> \right)^{-1/2}$  in Eq.~\eqref{eq:projectedSNR0}.

\begin{acknowledgments}
This work is supported by the Guangdong Provincial Key Laboratory (Grant No.2019B121203002),  the MURI-ARO Grant No. W911NF17-1-0323 through UC Santa Barbara, and the Shanghai Municipal Science and Technology Major Project through the Shanghai Research Center for Quantum Sciences (Grant No. 2019SHZDZX01). 
\end{acknowledgments}

\bibliography{apssamp}% Produces the bibliography via BibTeX.

\clearpage

\renewcommand{\thesection}{S\arabic{section}}
\renewcommand{\theequation}{S\arabic{equation}}
\setcounter{equation}{0}  %  this will re-count eq from 1
\renewcommand{\thefigure}{S\arabic{figure}}
\setcounter{figure}{0}  %  this will re-count eq from 1
\titleformat{\section} {\large\centering \bfseries }{\thesection.}{0.5em}{}
\titleformat{\subsection} {\centering \bfseries}{\thesubsection.}{0.5em}{}
\setcounter{page}{1}

\begin{widetext}
	
	\begin{center}
		\Large
		Supplementary Materials for     
	\end{center}
	
	\begin{center}
		\large 
		\bfseries
		Detecting axion dark matter with Rydberg atoms via induced electric dipole transitions
	\end{center}

	\begin{center}
		Georg Engelhardt$^* $,
		Amit Bhoonah $^\dagger$, and
		W. Vincent Liu$^\ddagger$
	\end{center}
	
	\begin{center}
		Corresponding authors:\\
		$^* $ engelhardt@sustech.edu.cn \\
		$^\dagger$ amit.bhoonah@pitt.edu \\
		$^\ddagger$ wvliu@pitt.edu
	\end{center}

	\textbf{ This PDF file includes:}

	\begin{itemize}
		\setlength{\itemsep}{0pt}
		\item[] Section \ref{sec:quantizationMaxwellEquations} to \ref{sec:sensitvity_superhet}. 
		\item[] Fig. \ref{figTemperatureDependence}.
		\item[] References
	\end{itemize}

	\setcounter{section}{0}
	\setcounter{subsection}{0}

	\section{Quantization of the axion-Maxwell equations}
	\label{sec:quantizationMaxwellEquations}
	
	In the article, we propose to deploy Rydberg atoms as highly sensitive dark matter detectors. In order to describe these quantum sensor in a consistent framework, we here perform a rigorous quantization of the axion-Maxwell equations. The structure of the section is as follows: In Sec.~\ref{sec:classicalAxionMaxwellEquations} we introduce the axion-Maxwell equation. In Sec.~\ref{sec:quantization}, we show how to accurately quantize these equations by microscopically deriving the Hamiltonian. In Sec.~\ref{sec:multiPolarHamiltonian}, we perform a Power-Zienau transformation, to bring it into a suitable form to describe spectroscopic experiments. 
	
	\subsection{Classical axion-Maxwell equations}
	
	\label{sec:classicalAxionMaxwellEquations}
	
	The relativistic Lagrangian density  describing the dynamics of the pseudo-scalar axion field $a$  is given by~\cite{Wilczek1987}
	\begin{eqnarray}
	\mathcal L  &=& -\frac{1}{4\mu_0} F_{\alpha\beta} F^{\alpha\beta} - A_{\alpha }J^{\alpha} \nonumber  \\
	&+& \frac{1}{2c\hbar}\partial_{\alpha} a \partial ^{\alpha} a - \frac{1}{2}\frac{ m_a^2c^4}{c^3\hbar^3} a^2 - \frac{g_{a\gamma\gamma}}{4\mu_0} a F_{\alpha\beta} \tilde F^{\alpha\beta} ,
	\label{eq:lagrangeDensity}
	\end{eqnarray}
	where the covariant electromagnetic field tensor $F_{\alpha\nu} =  \partial_{\alpha} A_{\beta} - \partial_{\beta }A_{\alpha}$  can be expressed in terms of the covariant 4-vector potential  $A^{\alpha} =  \left(  \phi , \boldsymbol A \right)  $. The labels   $\alpha,\beta,\gamma, \delta \in\left\lbrace ct, x,y,z\right\rbrace$ represent space-time coordinates.  The covariant electric 4-current is given as  $J^{\alpha} =  \left( c\rho,  \boldsymbol J \right)  $. The dual tensor of the electromagnetic field is defined by $ \tilde F^{\alpha\beta} =\frac{1}{2} \epsilon^{\alpha\beta \gamma\delta} F_{\gamma\delta} $, where $\epsilon^{\alpha\beta\gamma\delta}$ is the dyadic tensor.   The electric and magnetic fields can be obtained via
	\begin{equation}
	\boldsymbol E =- \boldsymbol  \nabla \phi - \dot {\boldsymbol  A }  \quad \text{and } \quad \boldsymbol  B = \boldsymbol  \nabla \times\boldsymbol  A.
	\end{equation}
	Using the Euler-Lagrange equations and the Bianchi identity, we find the following relativistic equations of motion for the above-introduced fields
	\begin{eqnarray}
	\partial_{\alpha }F^{\alpha\beta}  &=&  J^{\beta} - g_{a\gamma \gamma} \tilde F^{\alpha\beta}\partial_\alpha a ,\nonumber \\
	\left(\frac{1}{2c\hbar} \partial_{\alpha} \partial^{\alpha} + \frac{ m_a^2c^4}{c^3\hbar^3}  \right) a  &=&  -\frac{g_{a\gamma \gamma}  } {4\mu_0}  F_{\alpha\beta}\tilde F^{\alpha\beta} , \nonumber \\
	\partial_{\alpha} \tilde F^{\alpha\beta} &=& 0,
	\end{eqnarray}
	which in terms of the electric, magnetic, and axion fields reads as
	\begin{eqnarray}
	\boldsymbol \nabla\cdot \boldsymbol  E &=& \frac{\rho}{\epsilon_0}  - cg_{a \gamma\gamma } \boldsymbol  B \cdot \boldsymbol \nabla a \label{eq:SM:elelctricGauss} ,\\
	\boldsymbol  \nabla\times\boldsymbol   B - \frac{\dot {\boldsymbol  E}}{c^2}    &=& \mu_0  \boldsymbol  J 
	+ \frac{ g_{a\gamma\gamma }}{c} \left(\boldsymbol  B  \dot a - \boldsymbol  E \times\boldsymbol  \nabla a \right) \label{eq:SM:ampere} , \\
	\boldsymbol  \nabla \cdot \boldsymbol  B  &=&  0   \label{eq:SM:magneticGauss},  \\
	\boldsymbol   \nabla \times \boldsymbol  E  + \dot {\boldsymbol  B }  &=&   0  \label{eq:SM:faraday}, \\
	\ddot a - c^2\boldsymbol \nabla^2 a+ \frac{m_a^2c^4}{\hbar^2} a   &=&  \hbar c^3\sqrt{\frac{\epsilon_0}{\mu_0}  } g_{a\gamma\gamma} \boldsymbol  E\cdot \boldsymbol   B    \label{eq:SM:axionEom}.
	\end{eqnarray}
	These are the axion-Maxwell equations in SI units,
	which reduce to the common Maxwell equations  for $g_{a\gamma\gamma}=0$.  The axion field is given in units of $\text{eV}$, while the axion-photon coupling is given in units of $\text{eV}^{-1}$.

	\subsection{Quantization}
	
	\label{sec:quantization}
	
	Quantization of the axion-Maxwell equations means finding the microscopic Hamiltonian and the canonical commutation relation of the operators. We first state the result and then prove its correctness by deriving the non-relativistic classical axion-Maxwell equations using the Heisenberg equations of motion. The derivation  is a generalization of the derivation of the common Maxwell equations for $g_{a\gamma\gamma}=0$, which can be found in, e.g., Ref.~\cite{Mukamel1995}.

	In the Coulomb gauge $\boldsymbol \nabla \cdot \boldsymbol A =0$, the canonically quantized axion-light-matter Hamiltonian reads as
	\begin{eqnarray}
	\hat  H &=& \frac 12 \int d\boldsymbol r \left[ \epsilon_0\hat{\boldsymbol E}_c^{\perp 2 }  +  \frac{1}{\mu_0}\hat{\boldsymbol B}^2   \right]  \nonumber \\
	&+& \int d\boldsymbol r \left[   - \sqrt{\frac{\epsilon_0}{\mu_0}  }  g_{a\gamma\gamma}a\hat{\boldsymbol E}_c \cdot \hat{\boldsymbol B} + \frac{1}{2\mu_0} \left( g_{a\gamma\gamma} \hat a \right)^2 \hat{\boldsymbol B}\cdot \hat{\boldsymbol B} \right]  \nonumber \\
	&+&\frac{1}{2} \int d\boldsymbol r \left[ \frac{\hat \pi^2 }{\hbar c^3}  +  \frac{1}{c\hbar} \boldsymbol\nabla \hat a \cdot  \boldsymbol \nabla \hat  a  + \frac{ m_a^2c^4}{c^3\hbar^3} \hat a^2\right] \nonumber  \\
	&+& \sum_{\eta}\left[  \frac{\hat {\boldsymbol p}^2_{\eta }}{2 m_{\eta }}  +  \frac{q_{i}^2}{2 m_{\eta } c^2} \hat {\boldsymbol A}^2(\hat{ \boldsymbol r}_{\eta}) \right] \nonumber  \\
	&+& \frac{1}{2} \int d\boldsymbol r \left[ \hat   \rho \hat \phi  + \boldsymbol J \cdot\hat {\boldsymbol A}  \right]
	\label{eq:hamiltonian},
	\end{eqnarray}
	where the canonical electric field operator has been distributed in the transversal and longitudinal fields  $\hat {\boldsymbol E}_c = \hat {\boldsymbol E}^{\perp}_c +\hat {\boldsymbol E}^{\parallel }_c $. The  transversal  contribution is defined by $\boldsymbol \nabla \cdot  \hat {\boldsymbol E}^{\perp}_c  =0$, while the longitudinal contribution is given by  $\hat {\boldsymbol E}^{\parallel }_c  = \hat {\boldsymbol E}_c - \hat {\boldsymbol E}^{\perp}_c $. The electric potential $\hat \phi $ is defined via $\hat {\boldsymbol E}^{\parallel }_c = -\boldsymbol \nabla \hat \phi $. The vector potential and the magnetic field operators are denoted by $\hat {\boldsymbol B}$ and $\hat {\boldsymbol A}$, respectively.  The axion field is denoted by $\hat a$, and $\hat \pi$ denotes its conjugated momentum.  Particles with charge $q_\eta$ and mass $m_\eta$ at positions $\hat {\boldsymbol  r}_\eta$ with momentum $\hat {\boldsymbol  p}_\eta$ are labeled by $\eta\in \left\lbrace 1,\dots, N_p\right\rbrace$.
	The coupling between light and matter reflects the minimal coupling principle.  The commutation relations of the operators are given below.  The relation between the canonical and  the physical (observable)  electric field operators $\hat {\boldsymbol E}$ is established  via the relation
	\begin{equation}
	\hat {\boldsymbol E}  = 	\hat{\boldsymbol E}_c - cg_{a\gamma\gamma} \hat a \hat{ \boldsymbol  B}.
	\label{eq:SM:rel:canonicalPhysical}
	\end{equation}
	The canonical operators for all other operators agree with the physical operators. The field operators are quantized as
	\begin{eqnarray}
	\hat \rho(\boldsymbol  r) &=& \sum_{\eta=1}^{N_p} q_{\eta} \delta \left( {\boldsymbol  r} - \hat{\boldsymbol  r}_{\eta}  \right) ,\\
	\hat {\boldsymbol  J }(\boldsymbol r) &=& \sum_{\eta=1}^{N_p} q_{\eta}\dot{\hat{ \boldsymbol {r}}}_{\eta} \delta \left( {\boldsymbol  r} -\hat{ \boldsymbol  r}_{\eta}  \right) , \\
	\hat {\boldsymbol A} (\boldsymbol r) &=& \sum_{\boldsymbol k,\lambda} {\boldsymbol e}_{\boldsymbol k,\lambda} \sqrt{ \frac{\hbar }{2\omega_{\boldsymbol k}\epsilon_0 V}} \left( \hat d_{\boldsymbol k ,\lambda}^\dagger  e^{i\boldsymbol k\cdot \boldsymbol r}  +  \hat d_{\boldsymbol k ,\lambda}  e^{-i\boldsymbol k\cdot \boldsymbol r}   \right),  \label{eq:SM:def:vectorPotential}   \\
	\hat{\boldsymbol E}_c^{\perp }(\boldsymbol r) 
	&=&  i\sum_{\boldsymbol k,\lambda} {\boldsymbol e}_{\boldsymbol k,\lambda} \sqrt{ \frac{\hbar\omega_{\boldsymbol k}   }{2 \epsilon_0V}} \left( \hat d_{\boldsymbol k ,\lambda}^\dagger  e^{i\boldsymbol k\cdot \boldsymbol r}  -  \hat d_{\boldsymbol k ,\lambda}  e^{-i\boldsymbol k\cdot \boldsymbol r}   \right) , \label{eq:SM:def:electricFieldPerp}   \\
	\hat{\boldsymbol E}_c^{\parallel}( \boldsymbol r)  &=& -\boldsymbol \nabla \hat \phi( \boldsymbol r)  \label{eq:SM:def:electricFieldPara}  ,\\
	\hat \phi( \boldsymbol r)  &=&  \frac{1}{4\pi\epsilon_0}\sum_{\eta=1}^{N_p}\frac{q_{\eta}}{\left| \boldsymbol r - \hat{\boldsymbol r}_{\eta} \right|}, \label{eq:SM:electricPotential}\\
	\hat {\boldsymbol B}(\boldsymbol r) &=&   i\sum_{\boldsymbol k,\lambda} \sqrt{ \frac{\hbar  }{2\omega_{\boldsymbol k}  \epsilon_0 V}}  \left( \boldsymbol k  \times {\boldsymbol e}_{\boldsymbol k,\lambda}   \right)   \left( \hat d_{\boldsymbol k ,\lambda}^\dagger   e^{i\boldsymbol k\cdot \boldsymbol r}  -  \hat d_{\boldsymbol k ,\lambda}   e^{-i\boldsymbol k\cdot \boldsymbol r}   \right).
	\label{eq:SM:def:magneticField}
	\end{eqnarray}
	The photonic operators $\hat d_{\boldsymbol k,\lambda}$  quantizing the electromagnetic field are labeled by wavevector $\boldsymbol k$ and polarization $\lambda \in\left\lbrace  \updownarrow  , \leftrightarrow \right\rbrace$. The quantization volume is denoted by $V$. The unit vectors $ {\boldsymbol e}_{\boldsymbol k,\lambda}$ describe the direction of polarization and are perpendicular to the wavevector, i.e., ${\boldsymbol e}_{\boldsymbol k,\lambda} \cdot \boldsymbol k =0 $. This fact  makes sure that the electric and magnetic Gauss equations in Eqs.~\eqref{eq:SM:elelctricGauss} and \eqref{eq:SM:magneticGauss}  are fulfilled. The frequencies of the photonic modes are given by $\omega_{\boldsymbol k}$.

	Using  the definition of the electromagnetic field operators in Eqs.~\eqref{eq:SM:def:vectorPotential}-\eqref{eq:SM:def:magneticField}, we can show that the canonical commutation relations are indeed fulfilled:
	\begin{eqnarray}
	\left[\hat a  (\boldsymbol r)  , \hat \pi(\boldsymbol r^\prime ) \right]&=&  i \hbar^2 c^3 \delta(\boldsymbol r -\boldsymbol r^\prime)  \nonumber, \\
	\left[\hat {A}_{\alpha} (\boldsymbol r)  ,\hat { B}_{\beta} (\boldsymbol r^\prime ) \right] &=&  0 \nonumber,  \\
	\left[\hat {A}_{\alpha} (\boldsymbol r)  ,\tilde { E}_{c,\beta}^{\perp} (\boldsymbol r^\prime ) \right]&=& -\frac{\hbar}{\epsilon_0} \cdot \delta_{\alpha,\beta}^{\perp}\left(\boldsymbol r-\boldsymbol r^\prime \right) \nonumber  ,\\
	\left[\hat {B}_{\alpha} (\boldsymbol r)  ,\tilde { E}_{c,\beta}^{\perp} (\boldsymbol r^\prime ) \right]	&=& \frac{\hbar}{\epsilon_0} \epsilon_{\alpha,\beta,\gamma}  \frac{d}{dx_{\gamma}}\delta\left(\boldsymbol r-\boldsymbol r^\prime \right),
	\label{eq:commutationRelations}
	\end{eqnarray}
	where $\alpha,\beta,\gamma \in\left\lbrace x,y,z\right\rbrace$.  All other commutation relations vanish. We refer to Ref.~\cite{Mukamel1995}  for the technical definition of $\delta_{\alpha,\beta}^{\perp}\left(\boldsymbol r \right) $.
	
	We are now in position to derive the non-relativistic axion-Maxwell equations. The electric and magnetic Gaussian equations  are fulfilled since
	\begin{eqnarray}
	\boldsymbol \nabla \cdot  \hat { \boldsymbol E}_c  &=&  \boldsymbol  \nabla \cdot \left( \hat { \boldsymbol E}^{\perp}_c + { \boldsymbol E}^{\parallel}_c  \right) 
	=  \frac{ \hat \rho}{\epsilon_0},
	\end{eqnarray}
	where we have used that $ \boldsymbol \nabla \cdot\hat { \boldsymbol E}^{\perp}_c  = 0 $ and $ \boldsymbol \nabla \cdot  { \boldsymbol E}_c^{\parallel}  =\hat  \rho/\epsilon_0 $ because of the parameterization in Eqs.~\eqref{eq:SM:def:electricFieldPerp} and \eqref{eq:SM:def:electricFieldPara}. Using now the relation of the canonical and physical electric fields in Eq.~\eqref{eq:SM:rel:canonicalPhysical} we obtain the electric Gauss equation. For the same reason, we also find $ \boldsymbol \nabla \cdot\hat { \boldsymbol B} = 0 $, i.e., the magnetic Gauss equation in Eq.~\eqref{eq:SM:magneticGauss}.  
	
	The Faraday and Ampere equations can be constructed by means of the Heisenberg equations of motion
	\begin{equation}
	\dot{ \hat O} = \frac{i}{\hbar}\left[\hat H, \hat O\right] .
	\end{equation}
	The derivation follows essentially  the same lines as the common Maxwell equations upon simply replacing $ \hat{\boldsymbol E}$  by $ \hat{\boldsymbol E}_c$.  
	Using the commutation relations in Eq.~\eqref{eq:commutationRelations}, we obtain
	\begin{eqnarray}
	\dot{ \hat {\boldsymbol B}} &=& -\boldsymbol \nabla\times \hat{\boldsymbol E}_c^{\perp }  + cg_{a\gamma\gamma}\boldsymbol  \nabla\times \left(\hat a \hat {\boldsymbol B} \right)  \nonumber ,\\
	\dot{ \hat{\boldsymbol E}}_c^{\parallel } &=&- \frac{1}{\epsilon_0} \boldsymbol J^{\parallel }\nonumber  ,\\
	\dot{ \hat{\boldsymbol E} }_c^{\perp }  &=&- \frac{1}{\epsilon_0} \boldsymbol J^{\perp } + c^2   \boldsymbol \nabla\times \hat {\boldsymbol B}  \nonumber  \\
	&&-  \boldsymbol \nabla\times\left[  c g_{a\gamma}\hat a \left(\hat{\boldsymbol E}_c - cg_{a\gamma\gamma} \hat a \hat{\boldsymbol B} \right)\right].
	\label{eq:axionModifiedMaxwell:trans}
	\end{eqnarray}
	In  the magnetic equation, we can replace $\hat{\boldsymbol E}_c^{\perp }  \rightarrow \hat{\boldsymbol E}_c$ as  $\boldsymbol \nabla \times  \hat{\boldsymbol E}_c^{\parallel }   =0  $ since $ \hat{\boldsymbol E}_c^{\parallel }  =-\boldsymbol \nabla \hat \phi $  for the electrostatic potential $\hat \phi$ defined in Eq.~\eqref{eq:SM:electricPotential}. Moreover, the notion and derivation of  $\boldsymbol J^{\perp }$ and $\boldsymbol J^{\parallel }$, that fulfill $\boldsymbol J = \boldsymbol J^{\parallel } + \boldsymbol J^{\perp } $, can be found in standard textbooks on quantum electrodynamics. We proceed  to combine  the longitudinal and transversal electric equations
	\begin{eqnarray}
	\dot{\hat{\boldsymbol E}}_c  &=& \dot{ \hat{\boldsymbol E}}_c^{\parallel } + \dot{ \hat{\boldsymbol E}}_c^{\perp } \nonumber \\
	&=&   - \frac{\boldsymbol J}{\epsilon_0}   + c^2 \boldsymbol \nabla\times \hat {\boldsymbol B}  - c g_{a\gamma\gamma}  \boldsymbol \nabla \times\left( \hat a \hat{\boldsymbol E}_c \right) \nonumber \\
	&+& c^2 g_{a\gamma\gamma}^2\hat a  \boldsymbol \nabla\times(\hat a \hat {\boldsymbol B} )    - c^2g_{a\gamma\gamma}^2 \left(\hat a \hat {\boldsymbol B}\right)\times \boldsymbol \nabla \hat a   .
	\label{eq:axionModifiedMaxwell:trans2}
	\end{eqnarray}
	Resolving the magnetic  equation in Eq.~\eqref{eq:axionModifiedMaxwell:trans} for $ cg_{a\gamma\gamma} \boldsymbol \nabla\times \left(\hat a \hat {\boldsymbol B} \right)$ and inserting into the electric equation Eq.~\eqref{eq:axionModifiedMaxwell:trans2}, we obtain
	\begin{eqnarray}
	\dot{\hat{\boldsymbol E}}_c  &=&  -  \frac{\boldsymbol J}{\epsilon_0}  + c^2 \boldsymbol \nabla\times \hat {\boldsymbol B}   -c  g_{a\gamma\gamma}   \boldsymbol \nabla \times\left( \hat a \hat{\boldsymbol E}_c \right) \nonumber \\
	&+ &c g_{a\gamma\gamma}\hat a\dot {\hat {\boldsymbol B}} +c g_{a\gamma\gamma}\hat a \boldsymbol \nabla \times {\boldsymbol E}_c     - c^2 g_{a\gamma\gamma}^2 \left(\hat a \hat {\boldsymbol B}\right)\times \boldsymbol \nabla \hat a  \nonumber  \\
	&=&  -  \frac{\boldsymbol J}{\epsilon_0}  + c^2 \boldsymbol \nabla\times \hat {\boldsymbol B}   + c^2 g_{a\gamma\gamma}\hat a\dot{ \hat {\boldsymbol B} } \nonumber \\ 
	&+& c g \left(\hat{\boldsymbol E}_c -cg_{a\gamma\gamma} \hat a \hat {\boldsymbol B} \right)\times  \boldsymbol \nabla \hat a   .
	\end{eqnarray}
	Using now the relation of the  canonical and physical electric field operators in  Eq.~\eqref{eq:SM:rel:canonicalPhysical}, we readily find
	the Faraday and Ampere equations in Eqs.~\eqref{eq:SM:faraday}, \eqref{eq:SM:ampere} and \eqref{eq:SM:axionEom}, respectively. 
	
	We continue to construct the Heisenberg equations of motion for the axion field, which read
	\begin{eqnarray}
	\dot a &=&  \pi \label{eq:rel:aDot}\\
	\dot \pi &=&  c^2 \boldsymbol \nabla^2 \hat a  - \frac{ m_a^2c^4}{\hbar^2} \hat a   + \hbar c^3 \sqrt{\frac{\epsilon_0}{\mu_0}  }  g_{a\gamma\gamma}\hat{\boldsymbol E}_c \cdot \hat{\boldsymbol B} - \hbar c^3 \frac{1}{\mu_0} g_{a\gamma\gamma}^2 \hat a  \hat{\boldsymbol B}\cdot \hat{\boldsymbol B} .
	\end{eqnarray}
	Upon replacing the canonical electric field by the physical one according to Eq.~\eqref{eq:SM:rel:canonicalPhysical}, we obtain
	\begin{equation}
	\dot \pi = c^2  \boldsymbol \nabla^2 \hat a  - \frac{ m_a^2c^4}{\hbar^2} \hat a   + \hbar c^3  \sqrt{\frac{\epsilon_0}{\mu_0}  }  g_{a\gamma\gamma}\hat{\boldsymbol E} \cdot \hat{\boldsymbol B} \label{eq:dotPi2}.
	\end{equation}
	Deriving Eq.~\eqref{eq:rel:aDot} with respect to time and inserting Eq.\eqref{eq:dotPi2}, we readily obtain the axion equation of motion in Eq.~\eqref{eq:SM:axionEom}.

	\subsection{Multi-polar Hamiltonian}
	
	\label{sec:multiPolarHamiltonian}
	
	The minimal-coupling Hamiltonian in Eq.~\eqref{eq:hamiltonian0} is inconvenient to deal with because of the vector potential  $ \hat{ \boldsymbol A}$ that appears quadratically. This causes more complicated expressions when calculating, e.g., the spectroscopic response of quantum systems. Moreover, the vector potential $\hat{\boldsymbol A}$ is not a physical observable. For this reason, we will bring the Hamiltonian into the so-called multi-polar form that contains only the electric and magnetic fields~\cite{Mukamel1995}. This is done by means of the Power-Zienau transformation, which we review here shortly~\cite{Power1957}. 
	To this end, we restrict our investigation to systems with only bounded charges, such as atoms, molecules, and similar quantum emitters. We will specify to atoms in the following for clarity. Accordingly, the summation over $\eta$ in Eq.~\eqref{eq:hamiltonian0} will be modified into a double summation over $i \in \left\lbrace1,\dots,N \right\rbrace$, labeling atoms,  and $j \in \left\lbrace 1,\dots,N_c \right\rbrace$, labeling the charges associated with a particular atom.  The center of mass of atom $i$ will be denoted by $\boldsymbol R_i$, while the position of the charges belonging to this atom is denoted by $\boldsymbol r_{i,j}$.
	
	The Power-Zienau transformation  is defined  by the unitary operator
	\begin{equation}
	\hat U  = \exp\left[ -i\frac{1}{\hbar} \int d \boldsymbol r^3 \hat {\boldsymbol P} (\boldsymbol r)  \cdot \hat {\boldsymbol A} (\boldsymbol r)\right],
	\end{equation}
	where
	\begin{eqnarray}
	\hat {\boldsymbol P} (\boldsymbol r)  &=&   \sum_{m,j}\hat {\boldsymbol n}_{i,j}(\boldsymbol r),\nonumber \\ 
	\hat {\boldsymbol n}_{i,j}(\boldsymbol r)  &=&  q_{i,j} \left( \hat {\boldsymbol r}_{i, j} - \boldsymbol R_{i} \right) \nonumber \\
	&\times & \int_{0}^{1} du  \;\delta \left[ \boldsymbol r - \boldsymbol R_i - u \left( \hat{\boldsymbol r}_{i, j}  - \boldsymbol R_{i} \right) \right]
	\end{eqnarray}
	denote the macroscopic polarization and the polarization generated by the particles $\eta=(i,j)$ carrying charge $q_{i,j}$, respectively.   The Power-Zienau transformation has the following effect on the system operators:
	\begin{eqnarray}
	\hat {\boldsymbol r}^{\prime}_{i,j}   &=& \hat {\boldsymbol r}_{i,j} ,\nonumber \\
	\hat {\boldsymbol p}^{\prime}_{i,j}   &=& \hat{\boldsymbol p}_{i,j} - \frac{q_{e}}{c} \hat {\boldsymbol A}\left( \hat {\boldsymbol  r}_{i,j}  \right) - \int  d \boldsymbol r^3  \hat {\boldsymbol n}_{i,j} (\boldsymbol r) \times \hat {\boldsymbol B}(\boldsymbol r),\nonumber \\
	\hat d_{\boldsymbol k,\lambda }^{\prime} &=& \hat d_{\boldsymbol k,\lambda } + \left(\frac{i}{\hbar} \right) \sqrt{ \frac{\hbar  }{2 \epsilon_0\omega_{\boldsymbol k}  }}    {\boldsymbol e}_{\boldsymbol k,\lambda} \cdot \hat {\boldsymbol P} (\boldsymbol k).
	\end{eqnarray}
	Consequently, the electromagnetic field operators become
	\begin{eqnarray}
	\hat {\boldsymbol A}^{\prime} (\boldsymbol r)  &=&  \hat {\boldsymbol A} (\boldsymbol r) ,\nonumber\\
	\hat {\boldsymbol B}^{\prime} (\boldsymbol r)  &=&  \hat {\boldsymbol B} (\boldsymbol r)  ,\nonumber\\
	\hat {\boldsymbol E}_c^{\perp \prime} (\boldsymbol r)  &=&   \hat {\boldsymbol E}_c^{\perp }  (\boldsymbol r)   + \frac{1}{\epsilon_0} {\boldsymbol P}^{\perp} \equiv \frac{1}{\epsilon_0} \hat {\boldsymbol D}^{\perp} (\boldsymbol r) , \nonumber\\
	\hat {\boldsymbol P}^{\prime } (\boldsymbol r)  &=&  \hat {\boldsymbol P} (\boldsymbol r).
	\end{eqnarray}
	Thus, only the transverse canonical electric field becomes modified and is shifted by the polarization. The transformed electric field is called the displacement field $ \hat {\boldsymbol D} (\boldsymbol r)$. Importantly,  the displacement field is quantized in terms of the new photonic operators $\hat d_{\boldsymbol k,\lambda}^\prime$:
	\begin{eqnarray}
	\hat {\boldsymbol D}^\perp (\boldsymbol r)  =  i \sum_{\boldsymbol k,\lambda} \sqrt{ \frac{\hbar\omega_{\boldsymbol k}\epsilon_0   }{2 V }}    \boldsymbol  e_{k,\lambda} \left[\hat d_{\boldsymbol k,\lambda}^\prime    e^{i\boldsymbol k\cdot \boldsymbol  r} - \hat d_{\boldsymbol k,\lambda}^{\prime\dagger}   e^{-i \boldsymbol k\cdot\boldsymbol r}   \right] ,\nonumber\\
	\label{eq:SM:quantized_displacementField}
	\end{eqnarray}
	and not the canonical electric field $\hat {\boldsymbol E}_c (\boldsymbol r) $.  However, one should keep in mind that the electric field $\hat {\boldsymbol E} $ is the actual physical observable.   Away from the matter where $ \hat {\boldsymbol P} (\boldsymbol r) =0$, the displacement field is equivalent to the canonical electromagnetic field $\hat {\boldsymbol D} (\boldsymbol r)=\epsilon_0\hat {\boldsymbol E}_c (\boldsymbol r)$. In the absence of free charges, the displacement field is entirely transverse  and reads as
	\begin{equation}
	\hat {\boldsymbol D}^{\perp } (\boldsymbol r) =\hat {\boldsymbol D} (\boldsymbol r)   =  \hat {\boldsymbol E}_c (\boldsymbol r)   + \frac{1}{\epsilon_0}\hat  {\boldsymbol P} 
	\end{equation}
	as the longitudinal polarization fulfills $  \hat {\boldsymbol P}_c^{\parallel} = \epsilon_0\hat {\boldsymbol E}_c^{\parallel} =0 $ in the absence of free charges as considered here. 
	
	The  multi-polar Hamiltonian reads as
	\begin{equation}
	\hat H_{\text{ALM}} = \hat H_{\text{M}} + \hat H_{\text{L}} + \hat H_{\text{LM}} + \hat  H_{\text{A}} +  \hat H_{\text{Int}},
	\label{eq:multipoleHamiltonian}
	\end{equation}
	where the transformed matter Hamiltonian is given by
	\begin{eqnarray}
	\hat H_{\text{M}} &=&  \sum_{i,j} \frac{\hat p_{ij}^2 }{2 m_{ij}  } + \frac{1}{2\epsilon_0}\sum_{i,j,i',j'} \frac{q_{ij} q_{i'j}}{\left|\hat {\boldsymbol r}_{ij} -\hat {\boldsymbol r}_{i^\prime j^\prime} \right| }  \nonumber \\
	&+&   \frac{1}{2\epsilon_0} \int d \boldsymbol r^3 \left|  \hat {\boldsymbol P}^{\perp}(\boldsymbol r )  \right|^2.
	\end{eqnarray}
	The last term represents the polarization self-energy. The electromagnetic field Hamiltonian is now given as
	\begin{eqnarray}
	\hat  H_{\text{L}}  &=& \frac{1}{2} \int d \boldsymbol r^3 \left[\frac{1}{\epsilon_0} \hat {\boldsymbol D}^{\perp 2}(\boldsymbol r)  + \frac{1}{\mu_0}\hat {\boldsymbol B}^2(\boldsymbol r)  \right] \nonumber \\
	&=& \sum_{\boldsymbol k,x} \hbar \omega_{\boldsymbol k} \left(\hat d_{\boldsymbol k \lambda}^{\prime\dagger}\hat d_{\boldsymbol k \lambda}^\prime  +\frac 12  \right),
	\end{eqnarray}
	where in the second equality, we have expressed it in terms of the new photonic operators.  The transformed light-matter coupling now reads as
	\begin{eqnarray}
	H_{\text{LM}} = &-&  \int  d \boldsymbol r^3  \hat  {\boldsymbol P}(\boldsymbol r )\cdot \hat  {\boldsymbol D}^{\perp}(\boldsymbol r )  -\int  d \boldsymbol r^3  \hat  {\boldsymbol M}(\boldsymbol r )\cdot \hat  {\boldsymbol B}(\boldsymbol r ) \nonumber \\
	&+& \sum_{ij} \frac{1}{2m_{ij} c^2} \left[\int  d \boldsymbol r^3   \left[\hat n_{ij}(\boldsymbol r ) \times \hat B (\boldsymbol r)\right] \right] ^2.
	\end{eqnarray}
	The first term describes the coupling of the electric dipole moment to the displacement field. The second term is the coupling of the magnetic dipole moment (thoroughly defined in  Ref.~\cite{Mukamel1995}) to the magnetic field. The third term describes the coupling of the electric dipole density to the magnetic field. Yet, as it is divided by the rest energy of the charges $m_{ij} c^2$, this term can be safely neglected.
	
	The free axion Hamiltonian
	\begin{equation}
	\hat H_A = \frac{1}{2} \int d\boldsymbol r \left[ \frac{\hat \pi^2 }{\hbar c^3}  +  \frac{1}{c\hbar} \boldsymbol\nabla \hat a \cdot  \boldsymbol \nabla \hat  a  + \frac{ m_a^2c^4}{c^3\hbar^3} \hat a^2\right] \nonumber  \\
	\end{equation}
	remains unchanged, while the transformed axion-light-matter coupling Hamiltonian now reads as 
	\begin{eqnarray}
	\hat H_{\text{Int}} &=& c g_{a\gamma\gamma} \int d \boldsymbol r^3  \; \left[ \hat a \left( \hat {\boldsymbol D}   -  \hat {\boldsymbol P}  \right)  \cdot \hat  {\boldsymbol B}\right] \nonumber \\
	&+& \int d \boldsymbol r^3  \; \left[  \frac{1}{2\mu_0} g_{a\gamma}^2 \hat a^2 \hat{\boldsymbol B}\cdot \hat{\boldsymbol B} \right]
	\label{eq:axionCoupling}, 
	\end{eqnarray}
	where we have used that $c^2 = 1/ (\epsilon_0 \mu_0) $.

	\section{Derivation of the effective Hamiltonian}

	\subsection{Decoupling transformation}

	To simplify the following analysis, we consider a strong static magnetic field ${\boldsymbol B} $, which is in agreement with  our experimental proposal. The total magnetic field thus reads as
	\begin{equation}
	\hat {\boldsymbol B}  ={\boldsymbol B} +\hat {\boldsymbol B}_f ,
	\end{equation}
	where $\hat {\boldsymbol B}_f $ denotes the fluctuations of the field. We now carry out the unitary transformation
	\begin{equation}
	\hat U = \exp \left[ic g_{a\gamma\gamma}	\hat{\boldsymbol A} {\boldsymbol B}   \hat a  \right],
	\end{equation}
	which has the following effect on the system operators
	\begin{eqnarray}
	\hat{\boldsymbol D}  &\rightarrow&  	\hat{\boldsymbol D}  +  c g_{a\gamma\gamma}	 {\boldsymbol B}   \hat a ,\nonumber \\
	\hat \pi &\rightarrow & \hat \pi   +  c g_{a\gamma\gamma}	\hat{\boldsymbol A} {\boldsymbol B}  .
	\end{eqnarray}
	Thus, the transformed Hamiltonian now reads as
	\begin{eqnarray}
	\hat H&=&  \sum_{i,j} \frac{\hat p_{ij}^2 }{2 m_{ij}  } + \frac{1}{2\epsilon_0}\sum_{i,j,i',j'} \frac{q_{ij} q_{i'j}}{\left|\hat {\boldsymbol r}_{ij} -\hat {\boldsymbol r}_{i^\prime j^\prime} \right| }  \nonumber \\
	&+&   \frac{1}{2\epsilon_0} \int d \boldsymbol r^3 \left|  \hat {\boldsymbol P}^{\perp}(\boldsymbol r )  \right|^2.  \nonumber \\
	&+& \frac{1}{2} \int d \boldsymbol r^3 \left[\frac{1}{\epsilon_0} \hat {\boldsymbol D}^{\perp 2}(\boldsymbol r)  + \frac{1}{\mu_0} \left( {\boldsymbol B}(\boldsymbol r) +  \hat {\boldsymbol B}_f(\boldsymbol r) \right)^2  \right]  \nonumber \\
	&-&  \int  d \boldsymbol r^3 \frac{1}{\epsilon_0} \hat  {\boldsymbol P}(\boldsymbol r )\cdot \hat  {\boldsymbol D}^{\perp}(\boldsymbol r )  -\int  d \boldsymbol r^3  \hat  {\boldsymbol M}(\boldsymbol r )\cdot  \left( {\boldsymbol B}+  \hat {\boldsymbol B}_f \right)^2(\boldsymbol r) \nonumber \\
	&+& \sum_{ij} \frac{1}{2m_{ij} c^2} \left[\int  d \boldsymbol r^3   \left[\hat n_{ij}(\boldsymbol r ) \times \hat B (\boldsymbol r)\right] \right] ^2 \nonumber \\
	&-& c g_{a\gamma\gamma} \int d \boldsymbol r^3  \; \left[ \hat a \left( \hat {\boldsymbol D}   -  \hat {\boldsymbol P}  \right)  \cdot \hat  {\boldsymbol B_f}\right] \nonumber\\
	&+& \int d \boldsymbol r^3  \; \left[  \frac{1}{2\mu_0} g_{a\gamma}^2 \hat a^2 \left( {\boldsymbol B}+  \hat {\boldsymbol B}_f \right)^2(\boldsymbol r)^2 \right] \nonumber \\
	&+& \frac{1}{2} \int d\boldsymbol r \left[ \frac{\left(\hat \pi   +  c g_{a\gamma\gamma}	\hat{\boldsymbol A} {\boldsymbol B}   \right)^2 }{\hbar c^3}  +  \frac{1}{c\hbar} \boldsymbol\nabla \hat a \cdot  \boldsymbol \nabla \hat  a  + \frac{ m_a^2c^4}{c^3\hbar^3} \hat a^2\right] .
	\label{eq:SM:transformedHamiltonian}
	\end{eqnarray}
	In the transformed Hamiltonian, the axion-polarization coupling is mediated only by the fluctuations of the magnetic field $\hat  {\boldsymbol B_f}$, which can be thus neglected.
	As we see in the forth line, the polarization couples to the displacement field.
	For this reason, we have to determine its dynamics in order to estimate the axion field can induce some physically measurable effect on the polarization. The equations of motion of the displacement field reads as
	\begin{eqnarray}
	\frac{d}{dt}\hat {\boldsymbol D} &=&   - \frac{g_{a\gamma\gamma}}{\hbar c^2} \hat \pi(t) {\boldsymbol B} + c^2\boldsymbol \nabla \times {\boldsymbol B}_f  + c^2 \boldsymbol \nabla \times {\boldsymbol M} +\mathcal O \left(g_{a\gamma\gamma} \hat {\boldsymbol B}_f \right) \nonumber, \\
	\frac{d}{dt}\hat {\boldsymbol B}_f &=&  -\boldsymbol \nabla \times {\boldsymbol D} ,
	\label{eq:reducedMaxwellEquations}
	\end{eqnarray}
	which is a linearized version of the axion-Maxwell equations in Eqs.~\eqref{eq:SM:elelctricGauss} - \eqref{eq:SM:axionEom}.
	As $c^2$ is a large number, we have also constructed the equation of motion for the magnetic field fluctuations. For simplicity, we assume that the axion field is in a coherent state, such that its impulse has the following time evolution
	\begin{eqnarray}
	\hat \pi  \approx   m_a \dot a(t) \approx m_a a_0 \sin\left[m_a t \right],
	\end{eqnarray}
	i.e., we consider it as a classical source term. 
	
	\subsection{Solution of the axion-Maxwell equations}
	
	To solve the axion-Maxwell equations, we derive the first equation in Eq.~\eqref{eq:reducedMaxwellEquations}  with respect to time, and insert then  the magnetic field expression of the  second equation. Using identities of  vector calculus, we readily arrive at the inhomogenous wave equation
	\begin{eqnarray}
	\left[c^2\Delta -  \frac{d^2}{dt^2}  \right] {\hat {\boldsymbol D}}(r) &=&   - c g_{a\gamma\gamma} \dot{\hat \pi}(t) { \boldsymbol B}(r)  ,
	\label{eq:waveEquation}
	\end{eqnarray}
	where the axion field acts a source term. 
	This equation has the well-known solution
	\begin{eqnarray}
	{\hat {\boldsymbol D}}(\boldsymbol r,t)   &=&   - \frac{1}{4\pi }  \int d^3\boldsymbol r' \frac{  \boldsymbol f(\boldsymbol r' , t - \left|\boldsymbol r- \boldsymbol  r' \right|/c) }{\left|\boldsymbol r-  \boldsymbol r' \right| } ,
	\end{eqnarray}
	which can be derived, e.g., using the Green's function formalism~\cite{Economou2006}. In our case, the source term explicitly reads as
	\begin{eqnarray}
	\boldsymbol f(r, t)   &=&  c g_{a,\gamma\gamma} m_a^2 \cos(m_a t){ \boldsymbol B}(r). 
	\end{eqnarray}

	Without loss of generality, we  consider the position $\boldsymbol r =0 $ in the following, as  we can place the origin of the coordinate system at the position where we want to sense the displacement field.
	Expanding the magnetic field in terms of spherical harmonics $Y_{n,l}(\theta', \varphi')$,
	\begin{eqnarray}
	\boldsymbol B(\boldsymbol r') &=&  \boldsymbol B(r' ,\theta' ,\varphi')  
	= \sum_{n,l} \boldsymbol  B_{m,l}(r')Y_{m,l}(\theta', \varphi'),
	\end{eqnarray}
	we find  the following expansion for the displacement field
	\begin{eqnarray}
	{\hat {\boldsymbol D}}(\boldsymbol 0,t)   &=&  \sum_{l,m} {\hat {\boldsymbol D}}_{l,m}(\boldsymbol 0,t).
	\label{eq:displacementFieldExpansion}
	\end{eqnarray}
	The partial displacement fields can be expressed as
	\begin{eqnarray}
	{\hat {\boldsymbol D}}_{l.m}(0,t)   
	&=&   - \frac{ c  g_{a,\gamma\gamma}a_0 m_a^2}{4\pi }  \int_{0}^{\infty} dr' \int d\theta \int d\varphi  \boldsymbol B_{l,m}(r')Y_{l,m}(\theta', \varphi')   \cos(m_a( t -  r'/c) )   r' \sin \theta \nonumber \\ 
	&=&   - \frac{ c  g_{a,\gamma\gamma}a_0 m_a^2}{4\pi }  \mathcal K_{m,n} \int_{0}^{\infty} dr' \boldsymbol B_{l,m}(r')  \cos(m_a( t -  r'/c) )   r'  , 
	\label{eq:partialDisplacementField}
	\end{eqnarray}
	where the angular dependence is described by the coefficients
	\begin{equation}
	\mathcal K_{m,n}  =  \frac{1}{2\pi}  \int d\theta \int d\varphi  Y_{m,n} (\theta, \varphi)   \sin \theta .
	\end{equation}
	We emphasize that Eqs.~\eqref{eq:displacementFieldExpansion} and \eqref{eq:partialDisplacementField} comprise an exact solution of the inhomogeneous wave equation in Eq.~\eqref{eq:waveEquation}.

	For a practical reason, we want to bring Eq.~\eqref{eq:displacementFieldExpansion} into a more compact  form. To this end, we assume that all $\boldsymbol B_{l,m}(r')$ are parallel to the magnetic field at the origin $ { \boldsymbol B}(0)$. In this case, the displacement field can be expressed as
	\begin{equation}
	{\hat {\boldsymbol D}}(0,t)  =  \epsilon_0 c g_{a,\gamma\gamma} a_0 \cos(\omega_a t +\tilde \phi) { \boldsymbol B}(0) \mathcal S_{B}\left(m_a \right),
	\label{eq:displacementFieldTimeEvolution}
	\end{equation}
	where we have introduced
	\begin{eqnarray}
	\mathcal S_{B}\left(m_a \right)  &=& \left|  \mathcal F\right| \nonumber, \\ 
	\tilde \phi &=&  \arg \mathcal{ F} \nonumber , \\ 
	\mathcal F&=&   - \frac{1}{4\pi\left| { \boldsymbol B}(0)\right| }  \sum_{l,m} \mathcal K_{m,n} \int_{0}^{\infty} dr' \left| \boldsymbol B_{l,m}(r') \right| e^{-im_a r'/c }   r'  , 
	\end{eqnarray}
	which incorporates the form of the magnetic field. The phase $\tilde \phi$ will be neglected in the following.  It is interesting to analyze the form factor $F$ for small  masses $m_a$, such that $e^{-im_a r'/c }  \approx 1 $.  Rescaling  the argument of the magnetic field $ \boldsymbol B^\prime( r)  =\boldsymbol B( \alpha r) $, we find that $F^\prime \rightarrow \alpha^2 F$. We thus  conclude the following scaling behavior of the suppression factor
	\begin{equation}
	\mathcal S_{B}\left(m_a, R_B\right)  = \mathcal S \left( \frac{m_ac}{\hbar} 
	\cdot R_B \right) , 
	\label{eq:effectiveFormFunction}
	\end{equation}
	where $R_B$ parameterizes the spatial extend of the magnetic field, and the suppression function scales as $\mathcal S (x) \propto x^2$ for small $x$. The scaling for large $x$ depends on the particular form of the magnetic field.
	
	\subsection{Example }

	To illustrate the physical implications of  the general solution of the wave equation in Eq.~\eqref{eq:displacementFieldExpansion} on a concrete example, we consider an exponentially decaying magnetic field, 
	\begin{equation}
	\boldsymbol B_{l,m} (r'  ) = \boldsymbol B_{l,m;0} e^{-\gamma r'},
	\end{equation}
	where $\gamma =1/R_B$ denotes the inverse decay length. Inserting this into Eq.~\eqref{eq:partialDisplacementField} and evaluating the integral, we obtain
	\begin{eqnarray}
	{\hat {\boldsymbol D}}_{l,m}(0,t)  
	&=&   - c  g_{a,\gamma\gamma}a_0 m_a^2   e^{i m_a t}  \mathcal K_{m,n}\boldsymbol B_{l,m;0}    \int_{0}^{\infty} dr'   e^{-\gamma r'}   e^{-im_a r'/c}   r'   \nonumber  \\
	&=&   - c  g_{a,\gamma\gamma}a_0 m_a^2   e^{i m_a t}\mathcal K_{m,n} \boldsymbol B_{l,m;0}   \int_{0}^{\infty} dr'      e^{(-\gamma r'-im_a /c)r'}   r'  \nonumber  \\
	&=&    c  g_{a,\gamma\gamma}a_0 m_a^2   e^{i m_a t}\mathcal K_{m,n} \boldsymbol B_{l,m;0}   \int_{0}^{\infty} dr'    \frac{1}{-\gamma -im_a /c}  e^{(-\gamma -im_a /c)r'}    \nonumber  \\
	&=&   - c  g_{a,\gamma\gamma}a_0 m_a^2   e^{i m_a t} \mathcal K_{m,n} \boldsymbol B_{l,m;0}    \frac{1}{ \left(-\gamma -im_a /c \right)^2}  .   
	\end{eqnarray}
	The maximum amplitude of this field  has the following scaling properties:
	\begin{equation}
	{\hat {\boldsymbol D}}_{l,m}^{max} = c  g_{a,\gamma\gamma}a_0\boldsymbol B_{l,m;0}  \mathcal K_{m,n}  
	\begin{cases}
	m_a^2 R_B^2 & m_aR_B \ll 1 \\
	1 &  m_aR_B \gg 1
	\end{cases}   .
	\end{equation}
	For  small axion masses $m_aR_B \ll 1$, this relation shows that the axion-sourced displacement field is strongly suppressed, in agreement with Eq.~\eqref{eq:effectiveFormFunction}.  In contrast, for  $m_aR_B \gg 1$, the axion-sourced displacement field becomes independent of the axion mass and approaches the solution of the translationally  invariant system.

	\subsection{Effective Hamiltonian}
	
	Based on the previous derivations, we introduce here an effective Hamiltonian, which we use to predict the sensitivity of Rydberg atoms. First, we neglect the fluctuations of the electromagnetic fields and the axion field in the Hamiltonian in Eq.~\eqref{eq:SM:transformedHamiltonian}. In doing so, the Hamiltonian reduces to
	\begin{eqnarray}
	\hat H(t) &=&  \sum_{i,j} \frac{\hat p_{ij}^2 }{2 m_{ij}  } + \frac{1}{2\epsilon_0}\sum_{i,j,i',j'} \frac{q_{ij} q_{i'j}}{\left|\hat {\boldsymbol r}_{ij} -\hat {\boldsymbol r}_{i^\prime j^\prime} \right| }  
	+  \frac{1}{2\epsilon_0} \int d \boldsymbol r^3 \left|  \hat {\boldsymbol P}^{\perp}(\boldsymbol r )  \right|^2  \nonumber \\
	&-&  \int  d \boldsymbol r^3 \frac{1}{\epsilon_0} \hat  {\boldsymbol P}(\boldsymbol r )\cdot {\boldsymbol D}^{\perp}(\boldsymbol r,t )  -\int  d \boldsymbol r^3  \hat  {\boldsymbol M}(\boldsymbol r )\cdot   {\boldsymbol B}(\boldsymbol r) .
	\label{eq:simplifiedHamiltonian}
	\end{eqnarray}
	The magnetic field is considered to be static. The dynamics of the displacement field is given by Eq.~\eqref{eq:displacementFieldTimeEvolution}. Thereby, we implicitly assume  that all atoms are located closely to the origin as compared to the spatial variation of the magnetic field, i.e.,  $r_j\approx 0$. 
	
	Next, we represent the matter Hamiltonian in the energy basis of the atoms, such that it reads as
	\begin{equation}
	\hat H_0 = \sum_{i,j} \frac{\hat p_{ij}^2 }{2 m_{ij}  } + \frac{1}{2\epsilon_0}\sum_{i,j,i',j'} \frac{q_{ij} q_{i'j}}{\left|\hat {\boldsymbol r}_{ij} -\hat {\boldsymbol r}_{i^\prime j^\prime} \right| }    =  \sum_{i=1}^{N}\sum_{\mu,\nu }    \epsilon^{(i)}_{\mu} \left|i, \mu \right> \left< i, \mu \right|  , 
	\end{equation}
	where $ \epsilon^{(i)}_{\mu} $ denotes the energies. Likewise, we expand the magnetization operator
	\begin{equation}
	\hat {\boldsymbol M} (\boldsymbol r)   = \sum_{i=1}^{N}\sum_{\mu }   \boldsymbol  m^{(i)}_{\mu}  \left|i, \mu \right> \left< i, \mu \right|   \delta(\boldsymbol r -\boldsymbol r_i),
	\label{eq:ZeemanSM}
	\end{equation}
	and the polarization operator
	\begin{equation}
	\hat {\boldsymbol P} (\boldsymbol r)   = \sum_{i=1}^{N}\sum_{\mu,\nu }   \boldsymbol  d^{(i)}_{\mu,\nu}  \left|i, \mu \right> \left< i, \nu \right|   \delta(\boldsymbol r -\boldsymbol r_i),
	\label{eq:dipoleDensityOperatorSM}
	\end{equation}
	where $\delta(\boldsymbol r)$ is the three-dimensional delta function. Please note that the magnetization is assumed to diagonal here, such that it describes the common  Zeeman shift in atoms.  We assume that the atoms are far apart from each other, such that we can neglect the polarization interaction operator, i.e., the third term in Eq.~\eqref{eq:simplifiedHamiltonian}.
	
	After carrying out all these changes, the effective Hamiltonian   reads as
	\begin{equation}
	\hat H_{\text{eff}}(t) = \sum_{i=1}^{N}\sum_{\mu } \left[  \epsilon_\mu - \boldsymbol  m^{(i)}_{\mu} \cdot  {\boldsymbol B}_0 \right] \left|i, \mu \right> \left< i, \mu \right|      - \sum_{i=1}^{N}  \hat {\boldsymbol E}_{a}(t)  \cdot  \boldsymbol  d^{(i)}_{\mu,\nu}  \left|i, \mu \right> \left< i,\nu \right|,
	\label{eq:ham:axionDipole1SM}
	\end{equation}
	where the axion-sourced electric field is given by
	\begin{equation}
	\hat {\boldsymbol E}_{a}( t)  = g_{a\gamma\gamma}c  a(t)  {\boldsymbol B}_0 \cdot \mathcal S\left(  \frac{m_ac} {\hbar}\cdot R_B  \right)  .
	\label{eq:axionFieldElectricFieldRelationSM}
	\end{equation}
	Thereby, we have defined ${\boldsymbol B}_0 ={\boldsymbol B}(\boldsymbol r =0) $  as the magnetic field at the origin. The suppression factor, that depends on the axion mass $m_a$ and the spatial shape of the magnetic field has been defined in Eq.~\eqref{eq:effectiveFormFunction}. The second term in Eq.~\eqref{eq:ham:axionDipole1SM} is equivalent with Eq.~\eqref{eq:ham:axionDipole} in the article.

	\section{Sensitivity of  atomic superheterodyne detectors }
	
	\label{sec:sensitvity_superhet}
	
	In this section, we explain how to employ the celebrated electromagnetically-induced transparency (EIT) effect to construct an axion dark-matter detector featuring high sensitivity. The setup is called a superheterodyne (superhet) detector and has already demonstrated an outstanding sensitivity to weak electric fields~\cite{Jing2020}.  Thereby, we give a microscopic derivation of the projected signal-to-noise ratio (SNR)  given in Eq.~\ref{eq:projectedSNR0} in the article.
	
	\subsection{Hamiltonian }
	
	We describe the system as an effective four-level system whose states are denoted by $\left|1\right>$, $\left|2\right>$,  $\left|3\right>$, $\left|4\right>$. Their energetic positions are shown in the spectrum in Fig.~(2) in the article. The corresponding Hamiltonian reads as
	\begin{eqnarray}
	H(t) 
	&=& \epsilon_{1} \ketbra{1}{1}  +  \epsilon_{2} \ketbra{2}{2}  +  \epsilon_{3} \ketbra{3}{3}  + \epsilon_{4} \ketbra{4}{4}  \nonumber \\
	&+& \frac{\hbar}{2}\left[  \Omega_{p} e^{i\omega_p t}  \ketbra{2}{1} + h.c.  \right] \nonumber\\
	&+&\frac{\hbar}{2} \left[\Omega_{C} e^{i\omega_C t} \ketbra{3}{2} + h.c.  \right]\nonumber  \\
	&+& \frac{\hbar}{2}\left[ \left(  \Omega_{LO} e^{i\omega_{LO} t }  +  \Omega_a e^{i\omega_1 t }  \right)  \ketbra{4}{3} + h.c. \right],
	\label{eq:hamiltonian:superhet}
	\end{eqnarray}
	where $\epsilon_{x}$ with $x=1,2,3,4$ denote the level energies. The states $\ket{1}$ and $\ket{2}$ are coupled by the probe laser with frequency $\omega_p$ and Rabi frequency $\Omega_p$. The levels $\ket{2}$ and $\ket{3}$ are coupled by the coupling laser of frequency $\omega_C$ and Rabi frequency $\Omega_C$. The Rydberg states $\ket{3}$ and $\ket{4}$ are coupled via the axion-sourced electric field   of frequency $\omega_a$  and Rabi frequency $\Omega_a =g_{a\gamma\gamma}c  a_0 \left| \boldsymbol d_{3,4} \cdot    \boldsymbol B \right| /\hbar$, and via a  local oscillator field of frequency $\omega_{LO} $  and Rabi frequency $\Omega_{LO}  =  \left|  \boldsymbol  d_{3,4}\cdot \boldsymbol  E_{LO}^{(0)} \right|/\hbar $, where $\boldsymbol  E_{LO}^{(0)} $ is the amplitude of the corresponding electric field. The local oscillator $\Omega_{LO} \gg \Omega_a$ is deployed to linearize the measurement signal as a function of $\Omega_{a} $.
	
	In general, $\omega_a \neq \omega_{LO}$, but if $\left| \omega_a - \omega_{LO}\right|$ is smaller than all other frequencies in the system,  the two Rydberg states can be considered to be coupled by an effective field with frequency $\omega_{LO}$  and an adiabatically varying Rabi frequency $\Omega(t) = \Omega_{LO}  +  \Omega_a e^{i( \omega_a -\omega_{LO} ) t } $. To enable analytical calculations, we will consider $\Omega$ to be constant and parameterize it as $\Omega = \Omega_{LO} + \Omega_a$.

	To analytically describe the EIT in the four-level system which is subject to dissipation, we describe the dynamics  by the Bloch equation
	\begin{equation}
	\frac{d}{dt}\rho = -\frac{i}{\hbar} \left[ H(t) ,\rho\right]   + \gamma_2 D_{\ketbra{1}{2} } \rho   + \gamma_3 D_{\ketbra{1}{3}}+ \gamma_d D_{ \ketbra{1}{4}}    \rho,
	\label{eq:blochEquationLabFrame}
	\end{equation}
	where  the dissipator is defined by
	\begin{equation}
	D_ {\hat O  } \rho =  \hat O \rho \hat O^\dagger  - \frac{1}{2}\left( \hat O^\dagger \hat O \rho + \rho \hat O^\dagger \hat O \right) .
	\end{equation}
	and $\gamma_2$ , $\gamma_{3} $ and $\gamma_{4}$ denote the inverse life times of states $\ket{2}$, $\ket{3}$ and $\ket{4}$, respectively.
	
	For the ongoing analysis, we assume the resonance condition $\epsilon_{4} -\epsilon_{3} =\hbar \omega_a  $. Such a resonance condition can be fulfilled by adjusting the energies $\epsilon_{3},\epsilon_{4}  $ using the Zeemann effect. Transforming the subsystem consisting of the states $\epsilon_{3},\epsilon_{4}  $ into a frame rotating with $\omega_a$, and expressing the Hamiltonian using the diagonal basis of this subsystem, it becomes

	\begin{eqnarray}
	H(t) &=& \sum_{x=1,2,-,+}\epsilon_{x} \ketbra{x}{x} \nonumber \\
	&+& \frac{\hbar}{2}\left[  \Omega_{p} e^{i\omega_p t}  \ketbra{2}{1} 
	+ \frac{1}{\sqrt{2}}\Omega_{C} e^{i\omega_C t} \ketbra{-}{2} 
	+ \frac{1}{\sqrt{2}}\Omega_{C} e^{i\omega_C t} \ketbra{+}{2} 
	+ h.c. \right],
	\label{eq:ham:frame1}
	\end{eqnarray}
	where the energies in the diagonalized subspace are given by $\epsilon_{-} = \epsilon_{3} -\hbar\Omega/2$ and $\epsilon_{+} = \epsilon_{3} +\hbar\Omega/2$. Accordingly, the Bloch equation reads as
	\begin{equation}
	\frac{d}{dt}\rho = -\frac{i}{\hbar} \left[ H(t) , \rho \right]  + \gamma_2 D_{\ketbra{1}{2} } \rho   + \gamma D_{\ketbra{1}{-}} \rho + \gamma D_{ \ketbra{1}{+}}    \rho,
	\label{eq:blochEquations:frame1}
	\end{equation}
	where the dissipation rate $\gamma$ is  given by  $\gamma = \frac{1}{2} \left(\gamma_3 + \gamma_4 \right)$.
	
	\subsection{Signal operator}
	
	Using standard methods of spectroscopy~\cite{Mukamel1995}, one can show that the absorption rate of the probe laser is proportional to the imaginary part of the susceptibility $\text{Im}\, \chi(\omega_p)  $. As we demonstrate in the following, the susceptibility is intimately linked  to  the operator $\hat {\mathcal X} = \sum_{i=1}^{N} \hat {\mathcal X}_i$, where $i$ labels the $N$ atoms in the system, and
	\begin{eqnarray}
	\hat {\mathcal X}_i =\int_{0}^{t_{\text{m}}} dt \hat {\mathcal P}_{2,1 }^{(i)} (t) e^{ i(\omega_p t +\varphi)} + h.c. , \nonumber \\
	\hat {\mathcal P}_{2,1 }^{(i)} (t) \equiv    \hat U^\dagger(t)  \left( \left| 2 \right>_i\left< 1\right| +\left| 1 \right>_i\left< 2\right|  \right) \hat U(t) ,
	\label{eq:susceptibilityOperator}
	\end{eqnarray}
	where   $\hat U(t) $ is the time-evolution operator determined by the Hamiltonian in Eq.~\eqref{eq:hamiltonian:superhet}.  The phase $\varphi$ is a free parameter. The operator $\hat {\mathcal X}$ acts nonlocal in time, which is well defined in the Heisenberg picture. As we will show in the following, the expectation value for $\varphi=\pi/2$ and the variance for all $\varphi$ of $\hat {\mathcal X}$  are proportional to the imaginary part of the susceptibility in the linear response regime, i.e.,
	\begin{eqnarray}
	\left< \hat {\mathcal X}  \right>   \propto  \left< \hat {\mathcal X}^2 \right>   \propto  \text{Im}\, \chi(\Omega_p) , 
	\end{eqnarray}
	where the expectation value is defined by $\left< \bullet\right> =\text{tr} \left[\bullet \rho_{\text{st}}  \right]$ with the stationary state of the system for $\Omega_p =0$ given by $\rho_{\text{st}}  = \ketbra{1}{1}$. 
	Due to its close relation to the susceptibility (and thus to the absorption rate), we refer to $\hat {\mathcal X}$ as the susceptibility operator in the following. In Sec.~\ref{sec:signal-to-noise_ratio}, we thus take the operator $\hat {\mathcal X}$ as the basis to determine the SNR of the experimental setup. In doing so, we accurately account for the projection noise, which is the only fundamentally limiting noise source in the experiment~\cite{Cox2018}. Other noises, like photon shot noise or detection noise are not considered here.

	\subsection{Susceptibility}
	
	The susceptibility can be expressed in terms of the  linear response function $S(\boldsymbol r,t)$, which determines the polarization in response to  an external probe electric field $\hat {\boldsymbol E}_{p}(r,t) = \boldsymbol E_{p,0} \cos( \boldsymbol k \boldsymbol r - \omega_p t)$~\cite{Mukamel1995}, i.e.,
	\begin{equation}
	\hat {\boldsymbol P} (\boldsymbol r)  = \epsilon_0 \int d^3\boldsymbol r \int_{0}^{t} S(\boldsymbol r-\boldsymbol r^\prime, t-t^\prime) \hat {\boldsymbol E}_{p}(r^\prime, t^\prime) dt^\prime.
	\end{equation}
	We assume here that the probe frequency is close to resonance with the transition between $\ket{1}$ and $\ket{2}$ in the Hamiltonian in Eq.~\eqref{eq:hamiltonian:superhet}. For this reason, it is sufficient to parameterize the polarization operator as
	\begin{equation}
	\hat {\boldsymbol P} (\boldsymbol r)   = \sum_{i=1}^{N}  \boldsymbol  d^{(i)}_{1,2}  \hat {\mathcal P}_{2,1 }^{(i)} \delta(\boldsymbol r -\boldsymbol r_i).
	\label{eq:dipoleDensityOperator}
	\end{equation}
	Applying standard first-order perturbation theory to the Hamiltonian in Eq.~\eqref{eq:hamiltonian:superhet}, we find that the linear response function is given by~\cite{Mukamel1995}
	\begin{equation}
	S(\boldsymbol r,t) = -i \rho_N  \frac{  \left|\boldsymbol d_{1,2}  \right| ^{2} }{\epsilon_0\hbar }\left<\left[\hat {\mathcal P}_{2,1 } (t),\hat {\mathcal P}_{2,1 } (0) \right] \right>_{\Omega_p=0}\cdot \delta(\boldsymbol r),
	\end{equation}
	where $\rho_N$ denotes the atom density.
	Here, we have assumed that all atoms are equal, such that we can neglect the superscript $i$. The time-evolution is thereby determined for $\Omega_p=0$.
	The linear susceptibility can be obtained from the linear response function via Fourier transformation
	\begin{eqnarray}
	\chi(\omega_p)  &=&  \int d^3\boldsymbol r \int_{0}^{\infty} S(r,t) e^{i\omega_p t}dt  \nonumber \\
	&=&   (-i) \rho_N  \frac{\left|\boldsymbol d_{1,2}  \right| ^{2}  }{\epsilon_0\hbar } \left[  C(\omega_p)    - C(-\omega)^*    \right] \approx   (-i) \rho_N  \frac{\left|\boldsymbol d_{1,2}  \right| ^{2}  }  {\epsilon_0\hbar } C(\omega_p) .
	\label{eq:susceptibilityFrequency0}
	\end{eqnarray}
	For later purpose, we have introduced the correlation function
	\begin{eqnarray}
	C(t,t') &=&  \left. \left<  \hat {\mathcal P}_{1,2} (t)  \hat {\mathcal P}_{1,2} (t')  \right>\right|_{\Omega_p =0} \nonumber \\ 
	&=&  \left. \left< \hat {\mathcal P}_{1,2}(t-t')  \hat {\mathcal P}_{1,2}(0)  \right>\right|_{\Omega_p=0}  \nonumber \\
	& = &C(t-t').
	\label{eq:correlationFunction}
	\end{eqnarray}
	Its Fourier transformation is defined via
	\begin{equation}
	C(\omega)  = \int_0^{\infty} C (t)   e^{i\omega t} dt.
	\end{equation}
	In the approximation in Eq.~\eqref{eq:susceptibilityFrequency0} we have deployed the Kubo-Martin-Schwinger relation $C(\omega) =C(-\omega)e^{\beta \hbar \omega}  $ assuming room temperature $1/\beta \approx25 \text{meV}$ and probe frequency $\hbar \omega_p > 1 \text{eV}$.
	
	\subsection{Calculation of the signal}
	
	\label{sec:nmr:mean}
	
	We calculate the expectation value of the operator $
	\hat {\mathcal X}_i$ in Eq.~\eqref{eq:susceptibilityOperator} in the presence of the probe field $\hat {\boldsymbol E}_{p}(\boldsymbol r,t) = \boldsymbol E_{p,0} \cos( \boldsymbol k \boldsymbol r - \omega_p t)$ in   first-order perturbation theory  in  $\Omega_p = \left|\boldsymbol d_{1,2} \cdot \boldsymbol E_{p,0} \right| /\hbar   $.
	As trivially $\left< \hat {\mathcal X}_i \right>_{\Omega_p =0} =0  $, we find:
	\begin{eqnarray}
	\left< \hat {\mathcal X}_i \right> 	&\approx&    
	\Omega_p   \int_{0}^{t_{\text{m}}}dt  \int_{0}^{t}   dt' (-i)  \left[ C(t,t')   
	-  C(t',t) \right] \cos\left( \omega_p t' \right) \cos\left( \omega_p t  +\varphi \right)  .
	\end{eqnarray}
	In agreement with the RWA in the Hamiltonian in Eq.~\eqref{eq:hamiltonian:superhet}, we neglect the fast oscillating terms $e^{\pm 2\omega_p t}$. In doing so, we find
	\begin{eqnarray}
	\left< {\mathcal X}_i \right>  
	&\rightarrow&   \Omega_p  \int_{0}^{t_{\text{m}}}dt  \int_{0}^{t}   dt' (-i)\left[ C(t,t')   
	-  C(t',t) \right]  \left[ e^{i \left(\omega_p  t' -\omega_p t   + \varphi \right)}  +   e^{-i \left(\omega_p t' -\omega_p t + \varphi   \right)} \right] \nonumber \\
	& = &    \Omega_p \int_{0}^{t_{\text{m}}}dt  \int_{0}^{t}   dt' (-i) \left[ C(t')   
	-  C(-t') \right]  \left[ e^{i \left(-\omega_p  t'  + \varphi \right)}  +   e^{-i \left(-\omega_p t'  + \varphi   \right)} \right]   \nonumber   \\
	& = &\Omega_p   \int_{0}^{t_{\text{m}}}dt  \int_{0}^{\infty}   dt' (-i) \left[ C(t')   
	-  C(-t') \right]  \left[ e^{i \left(-\omega_p  t'  + \varphi \right)}  +   e^{-i \left(-\omega_p t'  + \varphi   \right)}  \right]\nonumber  \\
	& =  &   \Omega_p   \int_{0}^{t_{\text{m}}}dt    (-i)\left[    C(\omega_p) e^{ i \varphi  }   -  C(-\omega_p)^*  e^{-i \varphi   }  + C(-\omega_p) e^{- i \varphi  }   -  C(\omega_p)^*  e^{i \varphi   }  \right]  \nonumber  \\  
	& =  &    \Omega_p   t_{\text{m}}  (-i)\left[    C(\omega_p) e^{ i \varphi  }   -  C(-\omega_p)^*  e^{-i \varphi   }  + C(-\omega_p) e^{- i \varphi  }   -  C(\omega_p)^*  e^{i \varphi   }  \right]  .  
	\end{eqnarray}
	Setting $\varphi = \pi/2$ and comparing with Eq.~\eqref{eq:susceptibilityFrequency0}, we  obtain
	\begin{eqnarray}
	\text{Im} \,\chi(\omega_p)   & =  &   \frac{1}{2\Omega_p    t_{\text{m}}N }     \frac{\rho_N  \left|\boldsymbol d_{1,2}  \right| ^{2}    }{\epsilon_0\hbar } \left< \hat {\mathcal X}\right>  ,  
	\label{eq:signalSusceptibiliyRelation}
	\end{eqnarray}
	which establishes the desired relation between the operator $\hat {\mathcal X}$ and the susceptibility.

	\subsection{Calculation of the noise}
	
	\label{sec:nmr:noise}
	In the same manner, we evaluate the variance of the observable  $ \hat {\mathcal X} $. As $\Omega_p$  is assumed to be small, it has a minor influence on the result and we can set  $\Omega_p = 0$. Explicitly, the variance can be evaluated to be
	\begin{eqnarray}
	\left< \hat {\mathcal X}_i \hat {\mathcal X}_i  \right>  
	& = &    \int_{0}^{t_{\text{m}}}dt  \int_{0}^{t_{\text{m}}}   dt' \left< \hat {\mathcal X}_{1,2} (t)  \hat {\mathcal X}_{b,a} (t')  \right>_{\Omega_p =0}  \cos\left( \omega_p t'+\varphi  \right) \cos\left( \omega_p t  +\varphi \right)  \nonumber  \\
	& \rightarrow &    \frac{1}{2}    \int_{0}^{t_{\text{m}}}dt  \int_{0}^{t_{\text{m}}}   dt' C(t,t')  e^{i \left(\omega_p t' -\omega_p t   \right)}  \nonumber  \\
	&=&   \frac{1}{2}   \int_{0}^{t_{\text{m}}}dt  \int_{0}^{t_{\text{m}}}   dt'  C(t-t')  e^{i \left(\omega_p t' -\omega_p t   \right)}  \nonumber  \\
	&=&    \frac{1}{2}   \int_{0}^{t_{\text{m}}}dt  \int_{0}^{t}   dt'  C(t-t') e^{-i \left(\omega_p t' -\omega_p t  \right)}+
	\int_{0}^{t_{\text{m}}}dt  \int_{t}^{t_{\text{m}}}   dt' C(t'-t)^{*}   e^{-i \left(\Omega_p t' -\omega_p t   \right)}  \nonumber   \\
	&=&  \frac{1}{2}    \int_{0}^{t_{\text{m}}}dt  \int_{0}^{t}   dt'C(t')  e^{i \omega_p t'   }  +
	\int_{0}^{t_{\text{m}}}dt  \left[  \int_{0}^{t_{\text{m}}-t}   dt' C(t')    e^{i \omega_p t'    }   \right]^* \nonumber   \\
	&=&  \frac{1}{2}   \int_{0}^{t_{\text{m}}}dt  \int_{0}^{\infty}   dt'C(t')  e^{i \omega_p t'   }  +
	\int_{0}^{t_{\text{m}}}dt  \left[  \int_{0}^{\infty}   dt' C(t')  e^{i \omega_p t'    }  \right]^* \nonumber   \\
	&=&   \frac{1}{2}    t_{\text{m}} \left[ C(\omega_p)  +  C(\omega_p)^*  \right].
	%.
	%
	\label{eq:noiseSusceptibiliyRelation:trans}
	\end{eqnarray}
	Comparing this result with Eq.~\eqref{eq:susceptibilityFrequency0}, we infer that
	\begin{eqnarray}
	\text{Im} \,\chi(\omega_p)   & =  &     \frac{1}{    t_{\text{m}}  N }  \frac{\rho_N \left|\boldsymbol d_{1,2}  \right| ^{2}    }{\epsilon_0\hbar } \left< \hat {\mathcal X}^2 \right> ,
	\label{eq:noiseSusceptibiliyRelation}
	\end{eqnarray}
	where we have again deployed the Kubo-Martin-Schwinger relation.

	\subsection{Electromagnetically-induced transparency in a  four-level system }

	In  Secs.~\ref{sec:nmr:mean} and \ref{sec:nmr:noise}, we have derived formal expressions for the signal and the noise in terms of the susceptibility. It remains to express the susceptibility in terms of the microscopic system parameters. We achieve this  by solving the Bloch equation in Eq.~\eqref{eq:blochEquationLabFrame} and finding an explicit expression for the matrix element $\rho_{12}(t)$.  To connect this to the susceptibility, we use the  relation in first-order perturbation theory
	\begin{eqnarray}
	\rho_{12}(t)   &=& \int_{0}^{t} (-i)\left< \left[ \ketbra{1}{2}(t),\ketbra{2}{1}(t^\prime )\right] \right>_{\Omega_p=0} \Omega_p e^{i\omega_p t^\prime} \nonumber \\
	&=& \int_{0}^{t} \tilde C (t -t^\prime ) \Omega_p e^{i\omega_p t^\prime},
	\end{eqnarray}
	where we have introduced the correlation function $\tilde C (t ) $ for a notation reason.
	Performing a (normalized) Fourier transformation, we find
	\begin{eqnarray}
	\overline \rho_{21}(\omega_p) &=& \lim_{\tau\rightarrow \infty}  \frac{1}{2\tau} \int_{-\tau}^\tau \rho_{12}(t) e^{-i\omega_p t}dt\nonumber \\
	&=& \Omega_p\tilde  C(\omega_p),
	\label{eq:relation:correlationFunctionDensityMatrix}
	\end{eqnarray}
	where normalization is necessary to avoid a divergence. We note that due to the Kubo-Martin-Schwinger relation and within the RWA approximation $\tilde  C(\omega_p) \approx   C(\omega_p)$.  We thus obtain $C(\omega_p)$ and consequently the susceptibility in Eq.~\eqref{eq:susceptibilityFrequency0} via 
	\begin{equation}
	C(\omega_p)  = \left. \frac{d}{d \Omega_p} \overline \rho_{21}(\omega_p)\right|_{\Omega_p = 0}.
	\label{eq:rel:correlatorCoherence}
	\end{equation}
	
	To find an expression for $\overline \rho_{21}$, we generalize the standard treatment of the EIT in three-level systems (see, e.g.,  textbook of Scully~\cite{Scully1997}) to the four-level system in Eq.~\eqref{eq:hamiltonian:superhet}. To start with, we transform the Hamiltonian   into a frame rotating with frequencies $\omega_p$ and $\omega_C$, such that  the resulting time-independent Hamiltonian reads as
	\begin{eqnarray}
	H &=&  { \epsilon_{1,\Delta} }\ketbra{1}{1} +  {\epsilon_{2,\Delta} } \ketbra{2}{2}+  { \epsilon_{-,\Delta}} \ketbra{-}{c_1}  + \epsilon_{+,\Delta} \ketbra{+}{+}   \nonumber \\
	&+& \hbar\frac{ \Omega_{p}}{2}\left[ \vphantom{\sum} \ketbra{2}{1}  +  \ketbra{1}{2} \right] \nonumber \\
	&+& \hbar \frac{\Omega_{C}}{2\sqrt{2}} \left[\vphantom{\sum}  \ketbra{-}{2}  +  \ketbra{b}{-} \right] \nonumber  \\
	&+& \hbar \frac{\Omega_{C}}{2\sqrt{2}} \left[\vphantom{\sum} \ketbra{+}{2}  + \ketbra{2}{+} \right] ,
	\end{eqnarray}
	where  the detuned  energies are defined by $ \epsilon_{1,\Delta} = \epsilon_1 $, $ \epsilon_{2,\Delta} = \epsilon_2 -\hbar \omega_p  $,  $ \epsilon_{-,\Delta} = \epsilon_- -\hbar \omega_p -\hbar \omega_C $, and $ \epsilon_{+,\Delta} = \epsilon_+ -\hbar \omega_p -\hbar  \omega_C $. The dissipation terms in the Bloch equation in  Eq.~\eqref{eq:blochEquations:frame1} remain unchanged. Explicitly, the Bloch equations for the density matrix $\tilde \rho$ in the rotating frame  read as
	\begin{eqnarray}
	\frac{d}{dt}   \tilde \rho_{11}   &=&    \gamma_b  \tilde \rho_{22}  + \gamma  \tilde \rho_{--}  + \gamma  \tilde \rho_{++}  - i \frac{\Omega_{p} }{2}  \tilde \rho_{21}  +  i \frac{\Omega_{p} }{2}  \tilde \rho_{12} , \nonumber \\
	\frac{d}{dt}   \tilde \rho_{12}    &=&  -\frac{\gamma_b }{2}\tilde \rho_{12}  -  \frac{i}{\hbar}\left( \epsilon_{1,\Delta} - \epsilon_{2,\Delta} \right)\tilde \rho_{12}   -i\frac{\Omega_{p}}{2} \tilde \rho_{22} +i\frac{\Omega_{p}}{2}\tilde \rho_{11}  + i\frac{\Omega_{C}}{\sqrt{8}} \tilde \rho_{1-}    + i\frac{\Omega_{C}}{\sqrt{8}} \tilde \rho_{1+}   ,  \nonumber  \\
	\frac{d}{dt}   \tilde \rho_{1-}    &=&  -\frac{\gamma}{2} \tilde \rho_{1-}  - \frac{i}{\hbar} \left( \epsilon_{1,\Delta}-\epsilon_{-,\Delta} \right) \tilde \rho_{1-}      - i\frac{\Omega_{p}}{2} \tilde \rho_{2-}+i\frac{\Omega_{C}}{\sqrt{8}} \tilde \rho_{12}  ,\nonumber    \\
	\frac{d}{dt}   \tilde \rho_{1+}    &=&  -\frac{\gamma}{2} \tilde \rho_{1+}  - \frac{i}{\hbar} \left( \epsilon_{1,\Delta}-\epsilon_{+,\Delta} \right) \tilde \rho_{1+}    - i\frac{\Omega_{p}}{2}\tilde \rho_{2+}+i\frac{\Omega_{C}}{\sqrt{8}} \tilde \rho_{12}  ,  \nonumber  \\
	\frac{d}{dt}   \tilde \rho_{22}    &=&  -\gamma_2 \tilde \rho_{22} -i\frac{\Omega_{p}}{2} \tilde \rho_{12}  +i\frac{\Omega_{p}}{2}\tilde \rho_{21}       -i\frac{\Omega_{C}}{\sqrt{8}} \tilde \rho_{-2}  +i\frac{\Omega_{C}}{\sqrt{8}} \tilde \rho_{2-}  -i\frac{\Omega_{C}}{\sqrt{8}} \tilde \rho_{+2}  +i\frac{\Omega_{C}}{\sqrt{8}} \tilde \rho_{2+} , \nonumber   \\
	\frac{d}{dt}   \tilde \rho_{2-}    &=&   -\frac{1}2(\gamma_2+ \gamma) \tilde \rho_{2-}  - \frac{i}{\hbar} \left( \epsilon_{2,\Delta} - \epsilon_{-,\Delta} \right)\tilde \rho_{2-}  -i\frac{\Omega_{C}}{\sqrt{8}} \tilde \rho_{--}  -i\frac{\Omega_{C}}{\sqrt{8}} \tilde \rho_{+-}   +i\frac{\Omega_{C}}{\sqrt{8}} \tilde \rho_{22}  - i\frac{\Omega_{p}}{2}  \tilde \rho_{1-}, \nonumber  \\
	\frac{d}{dt}   \tilde \rho_{2+}    &=&   -\frac{1}2(\gamma_2+ \gamma) \tilde \rho_{2+}  - \frac{i}{\hbar}\left( \epsilon_{2,\Delta} - \epsilon_{-,\Delta} \right)\tilde \rho_{2+}  -i\frac{\Omega_{C}}{\sqrt{8}}\tilde \rho_{++}  -i\frac{\Omega_{C}}{\sqrt{8}} \tilde \rho_{-+}   +i\frac{\Omega_{C}}{\sqrt{8}} \tilde \rho_{22}  - i\frac{\Omega_{p}}{2}   \tilde \rho_{1+} ,\nonumber  \\
	\frac{d}{dt}   \tilde \rho_{--}    &=&  -\gamma \tilde \rho_{--}  -i\frac{\Omega_{C}}{\sqrt{8}} \tilde \rho_{2-}  +i\frac{\Omega_{C}}{\sqrt{8}} \tilde \rho_{-2} ,\nonumber  \\ 
	\frac{d}{dt}   \tilde \rho_{-+}    &=&   -\gamma \tilde \rho_{-+}  - \frac{i}{\hbar} \left( \epsilon_{2,\Delta} - \epsilon_{-,\Delta} \right)\tilde \rho_{-+}-i\frac{\Omega_{C}}{\sqrt{8}} \tilde \rho_{2+}  +i\frac{\Omega_{C}}{\sqrt{8}} \tilde \rho_{-2} ,\nonumber  \\
	\frac{d}{dt}   \tilde \rho_{++}    &=&  -\gamma \tilde \rho_{++}  -i\frac{\Omega_{C}}{\sqrt{8}} \tilde \rho_{2+}  +i\frac{\Omega_{C}}{\sqrt{8}} \tilde \rho_{+2}  .
	\end{eqnarray}

	We aim to find an expression for $   \tilde \rho_{12}$ in the leading order of $\Omega_p$. As inspection of the Bloch equations reveals that $\tilde \rho_{11}\approx 1\gg \tilde \rho_{22},\tilde \rho_{2-}, \tilde \rho_{-2}, \tilde \rho_{--}, \tilde \rho_{2+}, \tilde \rho_{-+}, \tilde \rho_{++}\propto \Omega_p^2$, we neglect corresponding terms.
	The three remaining equations thus are
	\begin{eqnarray}
	z \breve  \rho_{12}    &=&  -i\tilde \epsilon_{12} \breve  \rho_{12}   -i\frac{\Omega_{p}}{2} \breve \rho_{22}  + i\frac{\Omega_{C}}{\sqrt{8}}\breve  \rho_{1-}    + i\frac{\Omega_{C}}{\sqrt{8}}\breve  \rho_{1+}  +i\frac{\Omega_{p}}{2}\breve  \rho_{11}   ,\nonumber \\
	z  \breve  \rho_{1-}    &=&  -i\tilde \epsilon_{1-} \breve \rho_{12}     - i\frac{\Omega_{p}}{2}  \breve \rho_{2-}+i\frac{\Omega_{C}}{\sqrt{8}} \rho_{12}    \nonumber, \\
	z   \breve  \rho_{1+}    &=&  -i\tilde \epsilon_{1+}\breve  \rho_{12}  - i\frac{\Omega_{p}}{2} \breve  \rho_{2+}+i\frac{\Omega_{C}}{\sqrt{8}} \breve \rho_{12}   , 
	\label{eq:BlochEquationsLaplace}
	\end{eqnarray}
	which we have already transformed into Laplace space under the assumption that $\rho_{12}(0)= \rho_{1-}(0)= \rho_{1+}(0) =0$. Operators in Laplace space are marked by an inverted hat, i.e., $\breve \bullet$.  Moreover, we have defined
	\begin{eqnarray}
	\tilde \epsilon_{12}    &=&  \frac{1}{\hbar}(\epsilon_1 -\epsilon_2 )-\omega_p + i\frac{\gamma_2 }{2 },\nonumber \\
	\tilde \epsilon_{1-}   &=&  \frac{1}{\hbar}(\epsilon_1 -\epsilon_-) +\omega_p + \omega_C  + i\frac{\gamma }{2 },\nonumber \\
	\tilde \epsilon_{1+}    &=& \frac{1}{\hbar}( \epsilon_1 -\epsilon_+) +\omega_p + \omega_C  + i\frac{\gamma }{2 }
	\label{eq:parameterDefinition}
	\end{eqnarray}
	for a notation reason.  Resolving Eq.~\eqref{eq:BlochEquationsLaplace} for $\breve  \rho_{12}$, we obtain
	\begin{eqnarray}
	\breve  \rho_{12} = \frac{i\Omega_{p} \breve  \rho_{11}}{z +i \tilde \epsilon_{ab}  + \frac{1}{8}\frac{\Omega_{C}^2}{z +i \tilde \epsilon_{1-} }+\frac{1}{8} \frac{\Omega_{C}^2}{z +i \tilde \epsilon_{1+} }  }.
	\end{eqnarray}
	We are interested in the long-time behavior of $\tilde \rho_{12}(t\rightarrow \infty)$.
	Using $\rho_{11}(t) \approx \rho_{11}(0) = 1  $, which  in Laplace space becomes $\breve\rho_{11} = \rho_{11}(0) /z    $, we find
	\begin{eqnarray}
	\tilde \rho_{12}(t\rightarrow \infty) &=& \lim_{z\rightarrow 0} z  \breve\rho_{12}(z)  
	= \frac{-\Omega_{p}}{  \tilde \epsilon_{12} - \frac{1}{8} \frac{\Omega_{C}^2}{ \tilde \epsilon_{1-} }- \frac{1}{8} \frac{\Omega_{C}^2}{  \tilde \epsilon_{1+} }  } \nonumber \\
	&=& \overline \rho_{12} (\omega_p)
	\end{eqnarray}
	To establish the second equality, we have transformed the density matrix into the lab frame via
	$	\rho_{12}(t\rightarrow\infty )  = \tilde \rho_{12}(t\rightarrow \infty)  e^{i\omega_p t} $,
	and carried out the scaled Fourier transformation in Eq.~\eqref{eq:relation:correlationFunctionDensityMatrix}.

	\subsection{Temperature dependence}
	
	Usually, the superhet detector is operated at a finite temperature (e.g., $T=300\,\text{K}$), which will deteriorate its sensitivity. Two effects have to be taken into account. Firstly, the dephasing rates of the excited states $\gamma_2,\gamma_3$, and $\gamma_4$ will increase. The temperature-dependent dephasing rates can be calculated using the Alkali-Rydberg-Calculator (ARC) package~\cite{Robertson2021}. Overall, this effect is relatively small. Secondly, the temperature-induced motion of the atoms will give rise to a Doppler shift concerning the probe and coupling beams. To mitigate this effect, the probe and coupling beams shall propagate in opposing directions, as sketched in Fig.~2(f) in the article. Consequently,  both laser frequencies are Doppler shifted as
	\begin{eqnarray}
	\omega_p \rightarrow \omega_p - \frac{2\pi}{\lambda_p }v ,\nonumber\\
	\omega_C \rightarrow  \omega_C + \frac{2\pi}{\lambda_C }v ,
	\end{eqnarray}
	where $\lambda_p=2\pi c/\omega_p$ ($\lambda_C=2\pi c/\omega_C$) is the wavelength of the probe (coupling) beam, and $v$ is the velocity of an atom parallel to the beams.
	
	The Doppler-averaged  density matrix  elements can be obtained by evaluating the integral
	\begin{eqnarray}
	\overline \rho_{12D}(\omega_p) &=& \int_{-\infty}^{\infty} \tilde \rho_{12}\left(\omega_p - \frac{2\pi}{\lambda_p }v   , \omega_C + \frac{2\pi}{\lambda_C }v   \right) \frac{1} {\sqrt{2\pi k_B T/m_R}  } e^{-\frac{m_R v^2}{2 k_B T}}dv \nonumber \\
	&=& \int_{-\infty}^{\infty}   \frac{-\Omega_{p}}{  \tilde \epsilon_{12}- \frac{2\pi}{\lambda_p }v  - \frac{1}{8} \frac{\Omega_{C}^2}{ \tilde \epsilon_{1-} }- \frac{1}{8} \frac{\Omega_{C}^2}{  \tilde \epsilon_{1+} }  }  \frac{1} {\sqrt{2\pi k_B T/m_R}  }e^{-\frac{m_R v^2}{2 k_B T}} dv  \nonumber  \\
	&=& \int_{-\infty}^{\infty}   \frac{c_{1} }{  c_{2}- v} \frac{1}{ \sqrt{2\pi  }\sigma } e^{-\frac{v^2}{2 \sigma^2}}dv ,
	\end{eqnarray}
	where $m_R$ denotes the mass of the Rydberg atoms, and $T$ is the temperature of the thermal cloud. For a notation reason, we have introduced the parameters
	\begin{eqnarray}
	c_{1} &=& -\frac{\Omega_p\lambda_p}{2\pi  }, \nonumber  \\
	c_{2} &=& \frac{\lambda_p} {2\pi}\left(   \tilde \epsilon_{12}  - \frac{1}{8} \frac{\Omega_{C}^2}{ \tilde \epsilon_{1-} }- \frac{1}{8} \frac{\Omega_{C}^2}{  \tilde \epsilon_{1+} } \right) \nonumber , \\
	\sigma &=&\sqrt{ \frac{k_B T}{m_R}}.
	\end{eqnarray}
	This integral can be  solved analytically. Using that  $\text{Im} \, c_{2} >0$, we can transform
	\begin{eqnarray}
	\overline \rho_{12D}(\omega_p) 
	&=&  \int_{-\infty}^{\infty} dv\int_{0}^{\infty} dx (-i)\frac{c_{1}}{ \sqrt{2\pi}\sigma  }   e^{i(c_{2} -v) x }    e^{-\frac{v^2}{2\sigma ^2}} \nonumber \\
	&=& (-i)\frac{c_{1}}{ \sqrt{2\pi }\sigma  } \sqrt{2\pi } \sigma  \int_{0}^{\infty}dx    e^{ -\frac{\sigma ^2}{2} x^2 +i c_{2} x} \nonumber \\
	&=& (-i)\frac{c_{1}}{ \sqrt{2\pi  }\sigma } \sqrt{2\pi } \sigma  e^{-\frac{ c_{2}^2}{2\sigma^2}}   \int_{0}^{\infty} dx    e^{ -\frac{\sigma^2}{2} x^2 +i c_{2} x +\frac{ c_{2}^2}{2\sigma^2} }  \nonumber  \\
	&=& (-i)\frac{c_{1}}{ \sqrt{2\pi  } \sigma} \sqrt{2\pi } \sigma e^{-\frac{ c_{2}^2}{2\sigma^2}}  \int_{0}^{\infty} dx    e^{ -\frac{\sigma^2}{2} \left( x -i\frac{ c_{2}}{\sigma^2}\right)^2 }  \nonumber  \\
	&=& (-i)\frac{c_{1}}{ \sqrt{2\pi  } \sigma } \sqrt{2\pi } \sigma e^{-\frac{ c_{2}^2}{2\sigma^2}}\int^{\infty}_{-i\frac{ c_{2}}{\sigma^2}} dx    e^{ -\frac{\sigma^2}{2} x^2 }  \nonumber  \\
	&=& (-i)\frac{c_{1}}{ \sqrt{2\pi  } \sigma } \sqrt{2\pi } \sigma e^{-\frac{ c_{2}^2}{2\sigma^2}} \frac{\sqrt{2}} {\sigma}  \int^{\infty}_{-i\frac{ c_{2}}{\sqrt{2}\sigma}  }  dx    e^{ - x^2 }   \nonumber\\
	&=& (-i)c_{1}  \frac{\sqrt{2}} {\sigma}  e^{-\frac{ c_{2}^2}{2\sigma^2}} \frac{\sqrt{\pi} }{2  }  \left[1 - \text{erf} \left( -i\frac{ c_{2}}{\sqrt{2}\sigma} \right) \right], 
	\end{eqnarray}
	where in the last step we have introduced the complex-valued error function $\text{erf}(z)$. We continue to bring this into a form that can be more conveniently analyzed. To this end, we expand the error function as a continued fraction
	\begin{eqnarray}
	\text{erf} \left( z \right) = 1- \frac{1}{\sqrt{\pi }} \frac{e^{-z^2} }{z +\frac{1}{2z + \frac{2}{z + \frac{3}{z +\dots}}} },
	\end{eqnarray}
	such that we can express the density-matrix element as
	\begin{eqnarray}
	\tilde \rho_{12D}(\omega_p) &=&    \frac{c_{1}}{c_{2} + i \sigma  \Gamma\left(-ic_{2}/(\sqrt{2}\sigma ) \right) }  ,
	\end{eqnarray}
	where we have defined the form function 
	\begin{eqnarray}
	\Gamma(z)  &=& \frac{\sqrt{2}}{2z + \frac{2}{z + \frac{3}{z +\dots}} }  
	= \frac{\sqrt{2}}{\sqrt{\pi}} \frac{e^{-z^2}}{1 - \text{erf} \left( z\right)    } -\sqrt{2}z.
	\end{eqnarray}
	As this function describes the impact of the Doppler effect on the susceptibility, we will refer to it as the Doppler-effect form function in the following.

	Putting now everything together, the imaginary part of the susceptibility is given by
	\begin{eqnarray}
	\text{Im } \chi(\omega_p)  &=&      \frac{\rho_N  \left|\boldsymbol d_{1,2}  \right| ^{2}     }  {\epsilon_0\hbar }\text{Im }\; \frac{c_{1}}{c_{2} + i \sigma  \Gamma\left(-ic_{2}/(\sqrt{2}\sigma )\right) }   ,
	\label{eq:SM:susceptibilityFrequencyDoppler}
	\end{eqnarray}
	which is proportional to the absorption rate $\alpha(\omega_p)$.
	We emphasize that this expression is non-perturbative for all parameters. Inspection of Eq.~\eqref{eq:SM:susceptibilityFrequencyDoppler}  shows that the Doppler effect enhances the dephasing rate $\gamma_2^\prime \rightarrow \gamma_2 +\frac{\omega_p\sigma}{c}  \Gamma\left(-ic_{2}/(\sqrt{2}\sigma )\right)$. We depict the absorption rate as a function of coupling laser frequency detuning $\Delta \omega_C = \omega_C -(\epsilon_3 -\epsilon_2)/\hbar $ for different temperatures in Fig.~\ref{figTemperatureDependence}(a). 
	
	Without local oscillator $\Omega_{LO}=0$, the absorption rate exhibits a broad single dip at $\Delta \omega_C =0$, which is the celebrated EIT. Since $\text{Im } \chi(\omega_p) \propto \gamma$, the EIT is not complete for $T=300\,\text{K}$, as $\gamma$ increases for increasing temperature. The width of the dip scales with $\propto 1/\gamma_2^\prime$ and thus decreases for larger temperatures.
	
	For a finite local oscillator strength $\Omega = 0.1\, \text{MHz}$, the two levels $\ket{3}$ and $\ket{4}$ mix, which leads to constructive interference at $\Delta \omega_C =0$ and destroys the EIT. This generates a peak in the absorption rate, whose width is proportional to $\gamma$. The hight of the peak sensitively depends on $\Omega$.

	\begin{figure}
		\includegraphics[width=\linewidth]{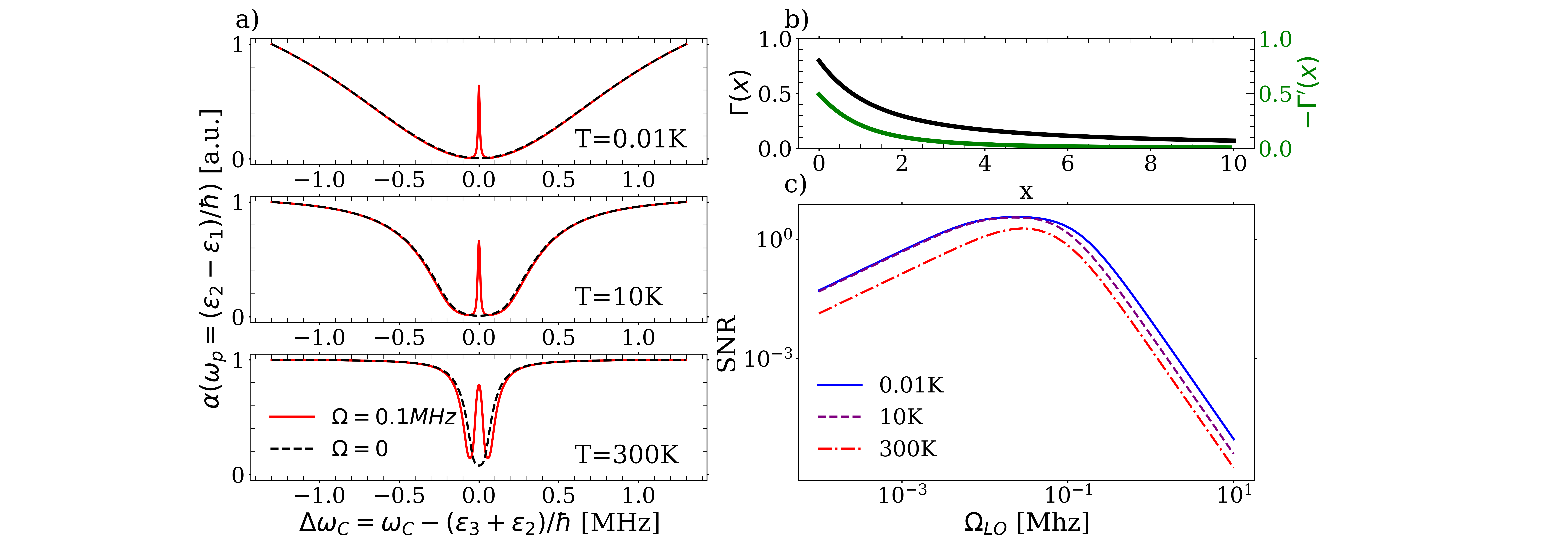}
		\caption{(a) Absorption rate as a function of $\Delta \omega_C$ for three different temperatures. (b) Doppler-effect form function $\Gamma(x)$ and its derivative $\Gamma^\prime (x)$. (c) SNR as a function of local oscillator strength for three different temperatures. Parameters are $\Omega_p =  1.7\,\text{MHz}$, $\Omega_C = 23\,\text{MHz}$, and  $B_{\text{res}} =5.6\,\text{T} $. If not specified differently, the parameters are the same as in Fig.~(2)(g). }
		\label{figTemperatureDependence}
	\end{figure}

	\subsection{Signal-to-noise ratio}
	
	\label{sec:signal-to-noise_ratio}
	As the susceptibility sensitively depends on $\Omega$ at the resonance condition $\Delta \omega_C=0$, the superhet detector operators under these circumstances, in which
	\begin{eqnarray}
	c_{2} =i \frac{\lambda_p} {2\pi} \left( \gamma_2  +  \frac{\Omega_C^2\gamma  }{\Omega^2 + \gamma^2 }  \right).
	\end{eqnarray}
	Based on Eqs.~\eqref{eq:signalSusceptibiliyRelation}, \eqref{eq:noiseSusceptibiliyRelation} and \eqref{eq:SM:susceptibilityFrequencyDoppler}, we can now evaluate the SNR for the operator $\hat {\mathcal X}$.
	Recalling that the total Rabi coupling between the Rydberg states consists of the local oscillator and the axion-induced dipole transitions $\Omega= \Omega_{LO} + \Omega_a$, we find that the signal for small $\Omega_a$ is given by
	\begin{eqnarray}
	\delta  \left< \hat {\mathcal X}\right>   &=&  \Omega_a \frac{d}{d\Omega_a} \left.\left< \hat {\mathcal X}\right>\right|_{\Omega_a =0}   \nonumber \\
	&=&  2\cdot     \frac{ 1    }{ \left(   \gamma_2 + \frac{\omega_p\sigma}{c}  \Gamma   + \frac{ \Omega_C^2\gamma}{ \Omega_{LO}^2 + \gamma^2  }  \right)^2   }    \left(  1 +  \Gamma^\prime \right) \frac{\Omega_C^2\gamma }{ \left(   \Omega_{LO}^2 + \gamma^2\right) ^2} 2 \Omega_{LO}  \cdot N  \Omega_p  \Omega_a t_{\text{m}}   ,
	\label{eq:signal:superhet}
	\end{eqnarray}
	where we have defined $\Gamma = \Gamma\left(\zeta / (\sqrt{2}\sigma)\right)$ with $\zeta =  \gamma_2+ \gamma\Omega_c^2/\left(\Omega_{LO}^2 + \gamma^2  \right)$ and  $\sigma^2 = k_b T/m_R$. Moreover, we have introduced  $\Gamma' = \partial_z \Gamma(z)\mid_{z\rightarrow\zeta / (\sqrt{2}\sigma) }$.
	Likewise, we can evaluate the variance for an ensemble of atoms
	\begin{eqnarray}
	\left<  \text{Var} \hat {\mathcal X}\right>  &=& \left< \hat {\mathcal X} ^2\right> 
	=      \frac{4 t_{\text{m}} N }{{\gamma_b}  + \frac{\omega_p\sigma}{c}  \Gamma  + \frac{\Omega_C^2\gamma  }{\Omega_{LO}^2 + \gamma^2 }     }  ,
	\end{eqnarray} 
	where we have approximated the  Rabi coupling between the Rydberg states by the local oscillator $\Omega\rightarrow \Omega_{L0}$. Putting everything together, the SNR  reads as
	\begin{eqnarray}
	SNR  &\equiv & \frac{\delta  \left< \hat {\mathcal X}\right>}{\left<  \text{Var} \hat {\mathcal X}\right>^{\frac{1}{2}}} \\
	&=&\sqrt{N} \Omega_a \sqrt{ T_c t_{\text{m} } }  =  \sqrt{N} \Omega_a \tau  ,
	\end{eqnarray}
	where we have defined the effective measurement time $\tau =\sqrt{T_c \cdot t_{\text{m}}}$ in terms of the effective coherence time
	\begin{eqnarray}
	T_c &=& \frac{4\Omega_p^2\Omega_C^4}{\left[ \left( \gamma_2+\frac{\omega_p\sigma}{c} \Gamma \right)  \left(\Omega_{LO}^2 + \gamma^2  \right)   + \Omega_C^2\gamma  \right]^3}\frac{ \gamma^2  \Omega_{LO}^2 }{\left( \Omega_{LO}^2 +\gamma^2 \right)  }\left(  1 +  \Gamma^\prime \right)^2 .
	\end{eqnarray}
	Finally, we recall that we have considered $\Omega$ as quasistatic throughout the derivations. Taking the time dependence $ \Omega(t) = \Omega_{LO}   +  \Omega_a e^{i( \omega_a -\omega_{LO} ) t } $ into account, we find that the measurement signal will slowly oscillate with frequency $\omega_a -\omega_{LO}$ and amplitude in Eq.~\eqref{eq:signal:superhet}. Thus, the SNR will remain unchanged in the case of heterodyning.

\end{widetext}

\end{document}